\documentclass[12pt]{article}
\usepackage{graphicx}
\usepackage{lineno}
\usepackage{hyperref}
\usepackage[usenames,dvipsnames]{color}
\usepackage{colordvi}
\newcommand{\captionfonts}{\small}
\makeatletter  
\long\def\@makecaption#1#2{%
  \vskip\abovecaptionskip
  \sbox\@tempboxa{{\captionfonts #1: #2}}%
  \ifdim \wd\@tempboxa >\hsize
    {\captionfonts #1: #2\par}
  \else
    \hbox to\hsize{\hfil\box\@tempboxa\hfil}%
  \fi
  \vskip\belowcaptionskip}
\makeatother    

\newcommand{\NIM}{{\it Nucl.~Instr.~Meth.~}}

\newcommand{\etal}{{\em et al.}}
\newcommand{\ind}{\hskip 0.25 in}

\begin{document}
\title{ 
{\bf Conceptual Design Report \\
A Compact Photon Source}}
\author{
%
\and\normalsize {B.~Wojtsekhowski} \\[-3pt]
\normalsize {\it Thomas Jefferson National Accelerator Facility, Newport News, VA 23606} 
\and\normalsize G.~Niculescu \\
[-3pt] \normalsize{\it James Madison University, Harrisonburg, VA 22807}
\bigskip
}
\date{June 22, 2015}
\maketitle
\tableofcontents

\bigskip
\begin{abstract} 
We propose to build a shielded magnet that will act as an untagged bremsstrahlung 
photon source suitable for deployment in Hall A/C at Jefferson Lab.
The goal of the project is a compact source which provides a narrow 1-mm diameter
photon beam at the target with the intensity adequate to the polarized target operation.
The proposed solution is based on the absorption of the electron beam in the magnet.
The compactness is achieved by means of a shallow channel which leads to an electromagnetic 
shower in the Cu-W body of the channel and surrounding material after a 1-2~mm deflection 
of the beam electrons by a horizontal magnetic field.
 
This source will allow us to achieve ten times the luminosity of the mixed electron--photon 
beam configuration while keeping the heat load on the (polarized) target as well as 
the radiation level in the experimental hall at acceptable levels.
The concept for this device was developed for the experimental proposal 
PR12-15-003 (\lq\lq{}WACS-ALL\rq\rq{}) but, if built, it would become 
a general--purpose source that would be usable for a variety of 
photon--induced reactions of interest in the 12-GeV era.

\end{abstract}

\bigskip
\section{Introduction}

\ind
The luminosity of an experiment is always a paramount parameter which needs to be considered.
In 12-GeV era experiments we often deal with large energy final state particles, so
the detector trigger rate-imposed limitations are not critical. 
At the same time, in the polarized target experiments the limit due to the polarized target heat
load limitation on the luminosity becomes a primarily parameter.
This is especially true for the polarized proton/deuteron target, which is based on super low
temperatures below 1$^\circ$K and loses its performance quickly with an increase in heat load.
The typical NH$_3$ target could operate at an average electron beam current of 90~nA or
a power deposition in the polarized material of 500 mW.

A large group of Jefferson Lab (JLab) 12-GeV era physics proposals (including PR12-15-003~\cite{WACS-ALL}) 
considers real-photon-induced processes. 
To study these reactions of interest, one needs a pure beam of real photons. 
At JLab, only Halls B and D have built-in real photon capabilities. 
This document presents a device which will act as a combination of a sweeper magnet 
and a beam dump, producing a narrow intense untagged pure photon beam. 
Because of its relatively modest size, the device can be deployed in either Hall A or Hall C  at JLab, 
thus opening the possibility of carrying out real-photon type experiments with so-called vertical 
bend spectrometers and making accurate measurements of the angular correlations.
There is also interest in such a source for Hall D due to the LOI12-15-001~\cite{Moskov}.

This document presents a preliminary CDR for the Compact Photon Source.
We plan to continue the analysis of every element of the device using MC simulation.
After approval of the physics proposal and completion of the CDR, we plan
to start the design and prototyping of the magnet insert (diffuser-absorber).

The following sections address the layout of the device, the TOSCA model and results
of magnetic analysis, and the GEANT4-based simulation code and its validation.
We also analyzed the radiation dose rate at key locations around the device: 
at the magnet coils and the polarized target solenoid, outside the shielding (1.25 m from the source), 
and in Hall A (an average of 15 m from the source). 
Distribution of the radiation in the forward direction is considered in detail. 
The heat load on the polarized target was found via an analytical estimate and from the 
MC of the full experimental setup.

\section{Magnet-Dump as a Compact Photon Source}
\label{sec:concept}

\ind A traditional source of bremsstrahlung photons includes a radiator, a deflection magnet with
large momentum acceptance and a dump for the used electron beam.
Such a configuration requires significant space and shielding.
In addition, it leads to a large size of the photon beam at the target due to
the natural divergence (~1/$\gamma$) of the photon beam.
To overcome these problems we have proposed the Compact Photon Source~\cite{WACS-ALL}. 

{\bf We propose to take advantage of the narrowness of the photon beam.
Without a loss of intensity, a narrow channel could be made around the photon beam.} 
Such a channel made of heavy metal (Copper-Tungsten alloy, a radiation length of a few mm) 
will serve as a diffuser for used electrons, an absorber for the shower, and 
a collimator of secondary particles produced in the electromagnetic and hadronic shower. 
There is a slow raster of the beam which moves the beam across the area of the polarized
target with a typical frequency of 50~Hz. 
In the proposed device it will be synchronized with the slow motion of the channel, and 
the electron beam will be turned off briefly (for a few $\mu$s) when 
the transition from one narrow channel to another is needed.

The principal components of the proposed device are listed below and will be discussed in detail 
in the remainder of this document:
\begin{itemize}
\item{A 1.3-mm copper radiator (10\% radiation length) located inside the shielded area.}
\item{A normal conducting dipole magnet (``amagnet") providing a strong horizontal field to sweep 
down the beam electrons.} 
\item{A 150-cm long W/Cu block with a set of 2-mm channels located in the dipole field 
which serves as a diffuser-absorber of an electromagnetic shower.
The diffuser-absorber movement, synchronized with the beam position raster, is
required for the polarized target.}
\item{A one meter thick layer of heavy radiation shielding to ensure that 
the radiation level in the Hall does not exceed allowable limits.}
\end{itemize}

\subsection{Magnet}

\begin{figure}[!h] 
\begin{minipage}[h]{0.48\textwidth}
\begin{center}
\includegraphics[trim = 30mm 30mm 20mm 20mm, width=6cm]
{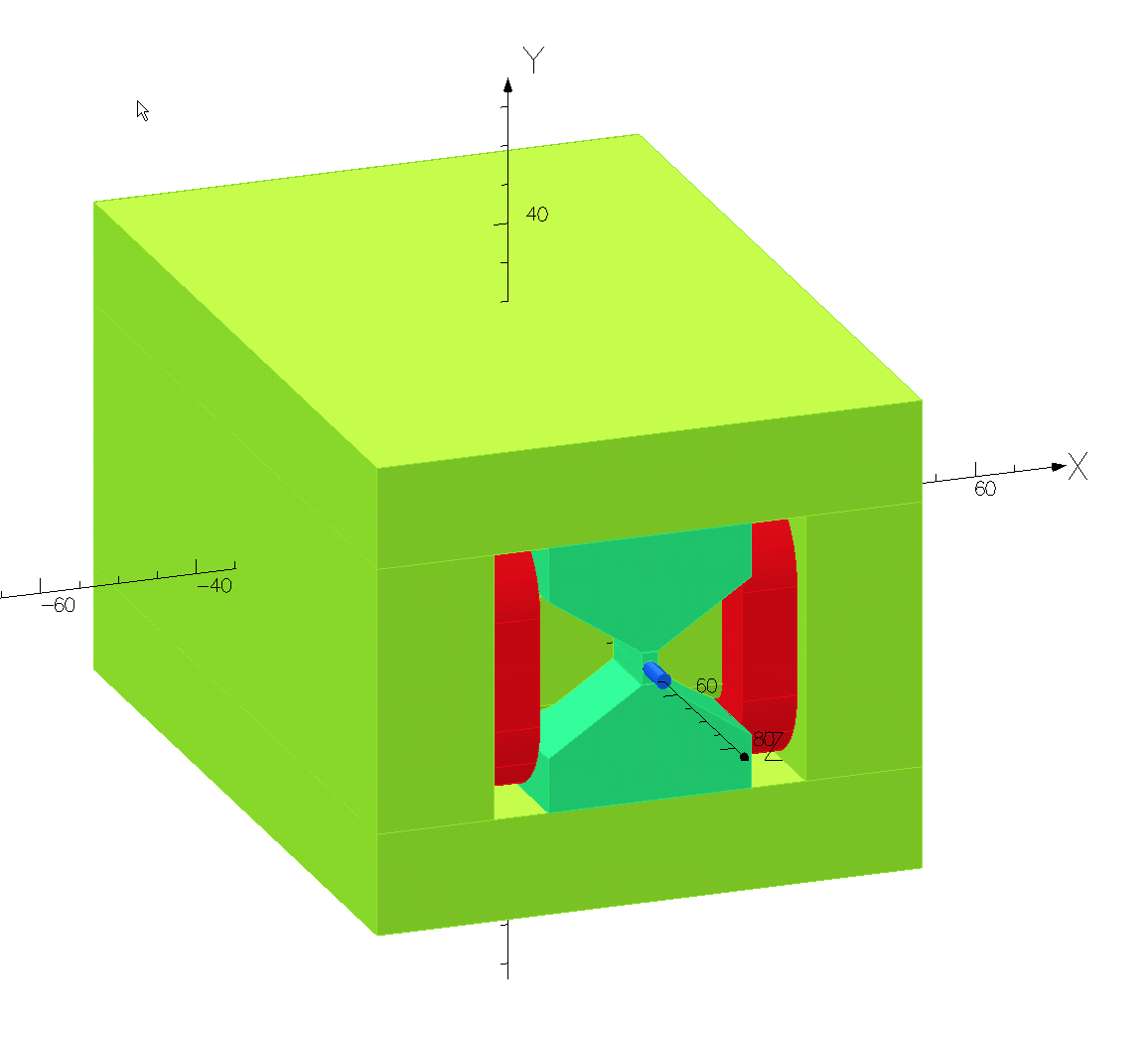}
\end{center}
\caption{Schematic of the sweeper magnet. 
The Cu inserts (blue--green) act as shielding in the immediate vicinity of the beam. 
The central area of the magnet bore is filled by a W/Cu diffuser-absorber.}
\label{fig:amagnet01a}
\end{minipage}
\hskip 0.5 in
\begin{minipage}[h]{0.45\textwidth}
\begin{center}
\includegraphics[trim = 30mm 30mm 30mm 0mm, width=4.cm]{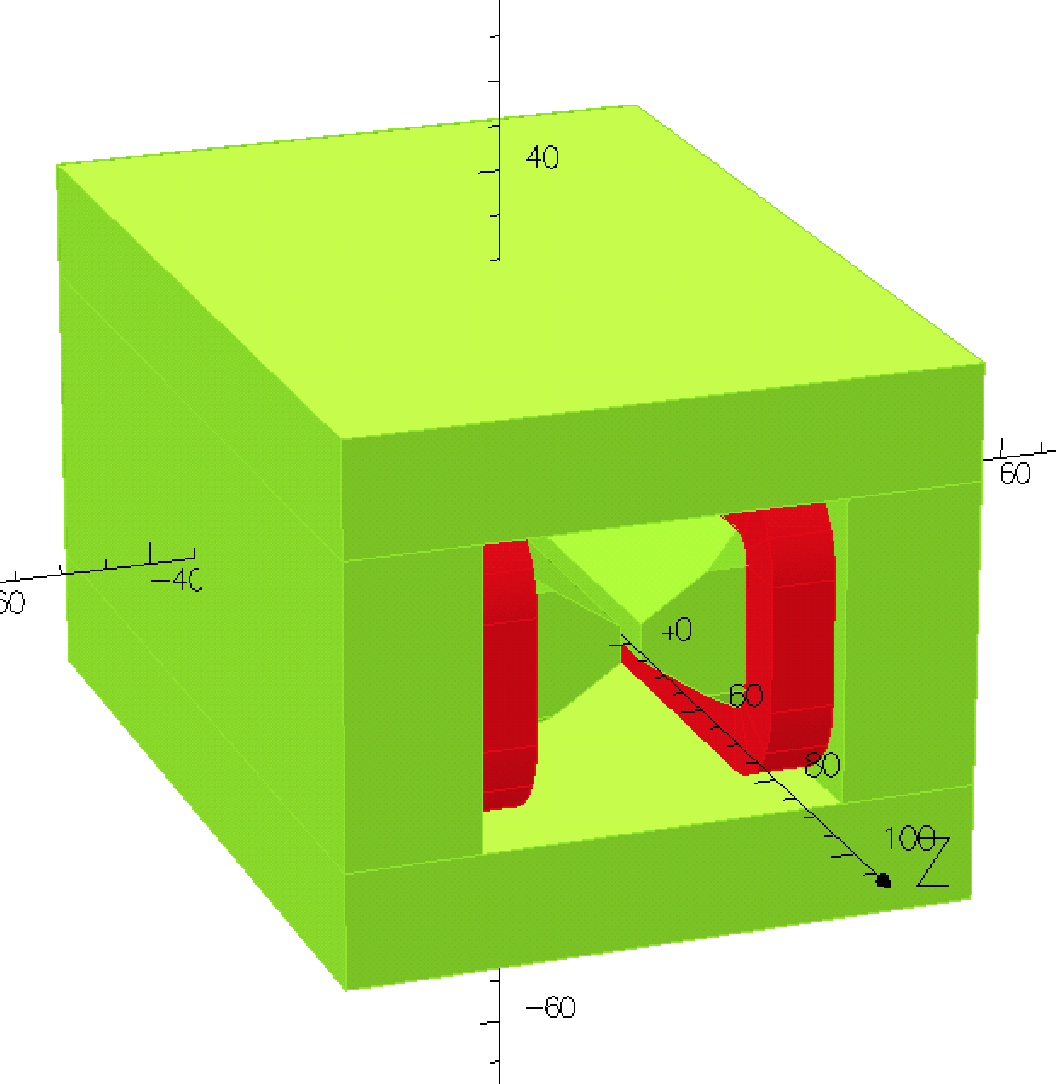}
\end{center}
\vskip 0.25 in
\caption{The sweeper magnet without 
the copper inserts.} 
\vskip 0.25 in
\label{fig:amagnet01b}
\end{minipage}
\end{figure}

\begin{figure}[!h]
\begin{minipage}[h]{0.50\textwidth}
\begin{center}
\includegraphics[trim = 30mm 30mm 20mm 20mm, width=5cm]{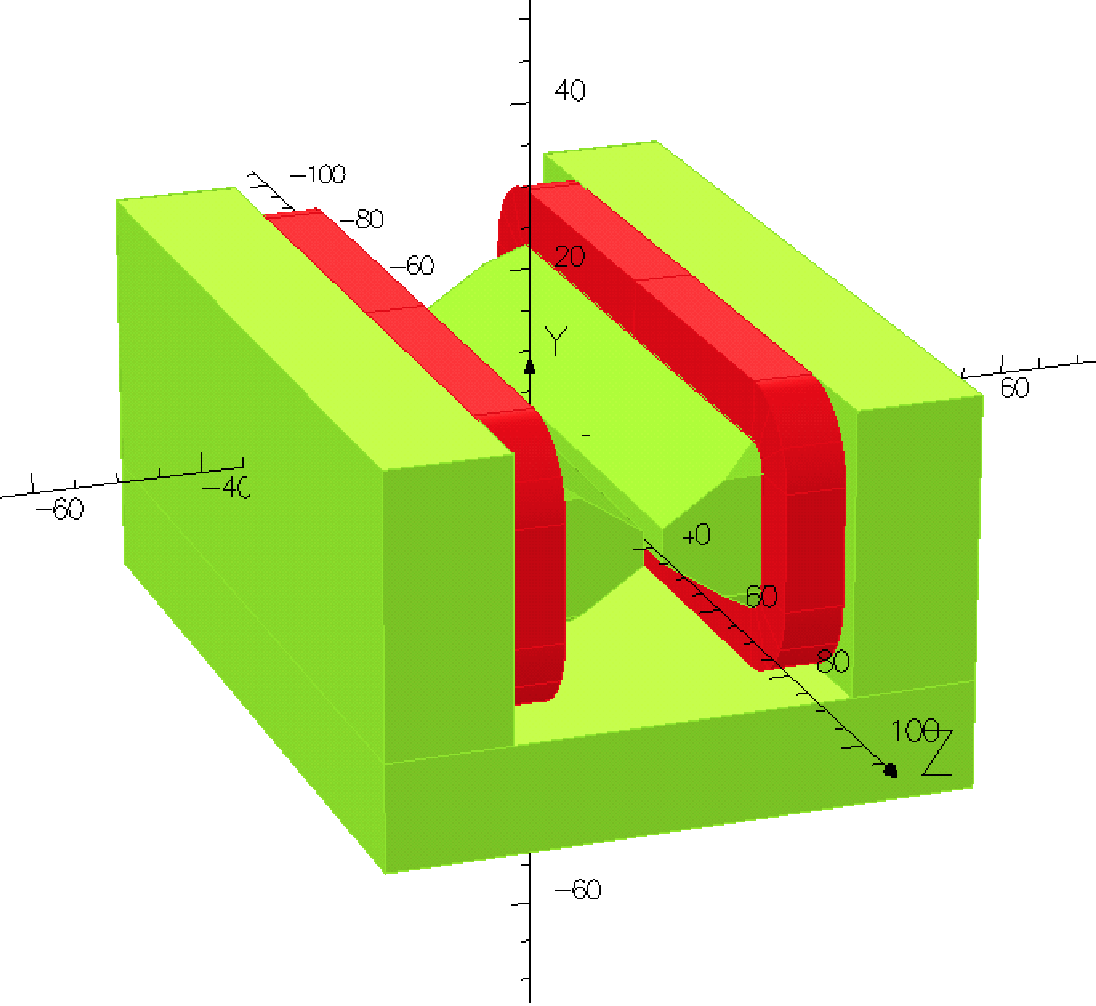}
\end{center}
\caption{The sweeper magnet without the top iron plate.} 
\label{fig:amagnet01c}
\end{minipage}
\begin{minipage}[h]{0.50\textwidth}
\begin{center}
\includegraphics[trim = 30mm 30mm 20mm 20mm, width=5cm]{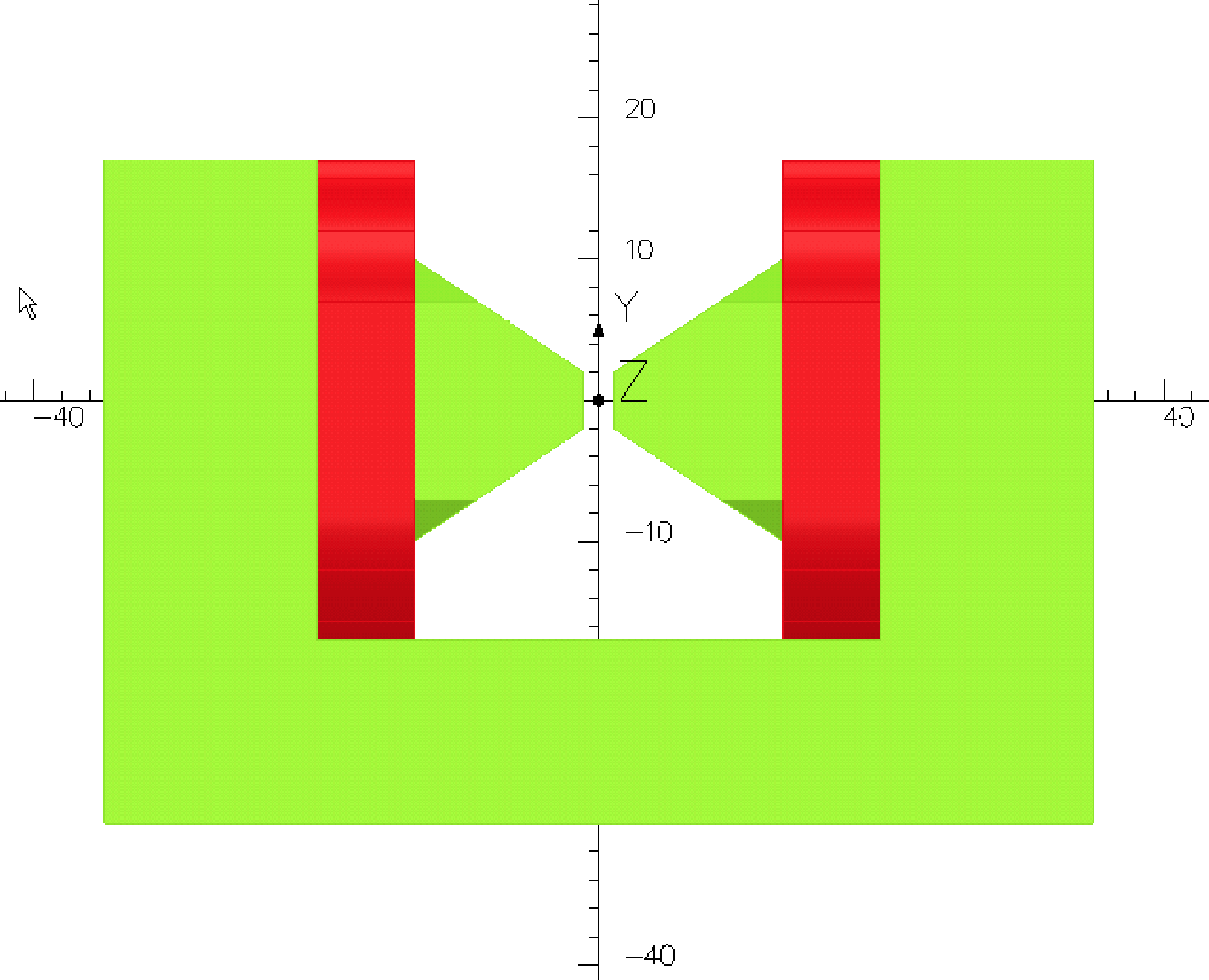}
\end{center}
\caption{The sweeper magnet front view.} 
\label{fig:amagnet01d}
\end{minipage}
\end{figure}

\begin{figure}[!h]
\begin{center}
\vskip -0.25 in
\includegraphics[trim = 0mm 0mm 0mm 0mm, width=8cm]{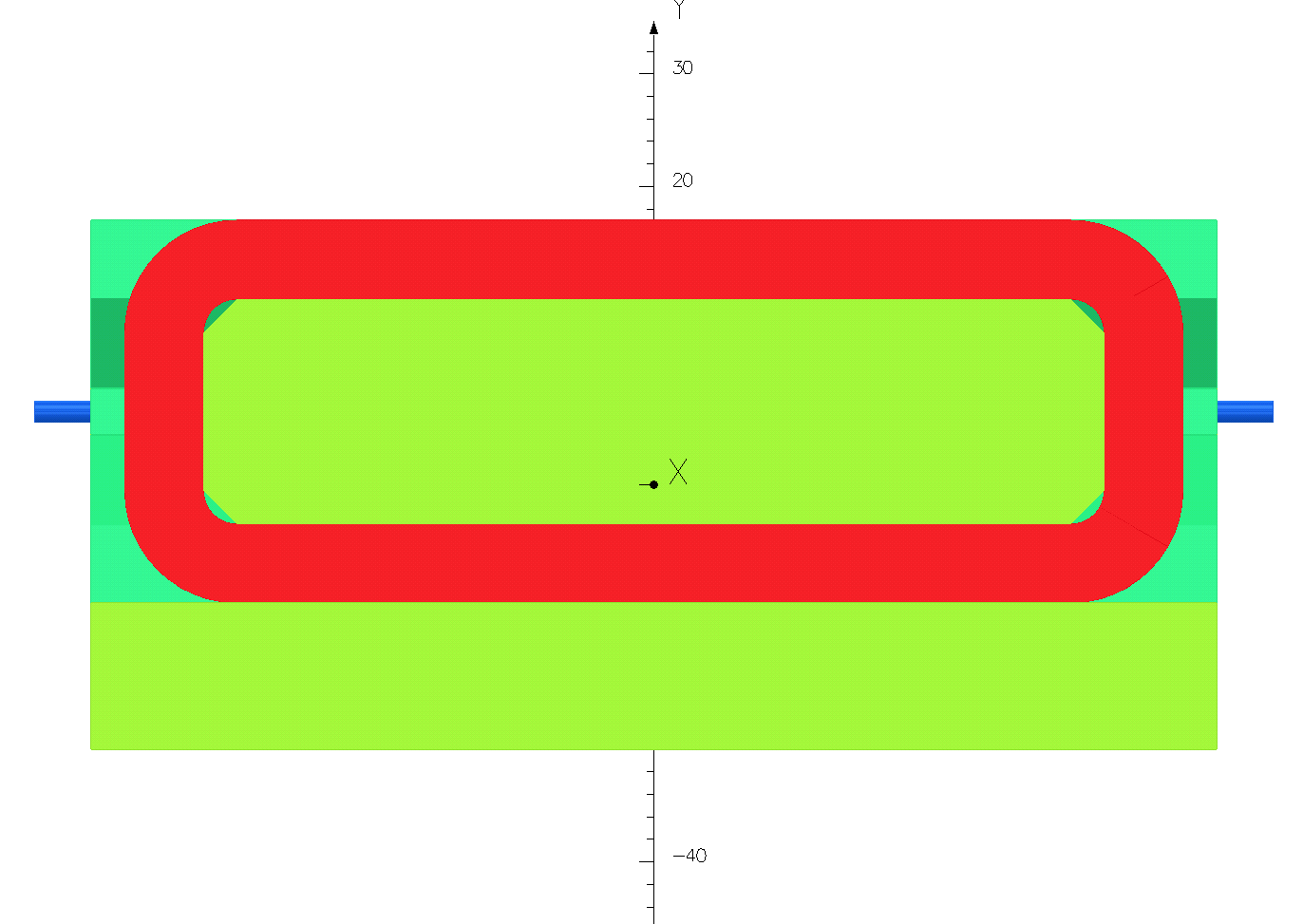}
\end{center}
\caption{The sweeper magnet side view (with the copper insert).} 
\label{fig:amagnet01e}
\end{figure}

Figures~\ref{fig:amagnet01a}-\ref{fig:amagnet01e} show the TOSCA model of 
the proposed magnet (100x60x50 cm$^3$) in several different projections partly 
open (some parts are removed) for clarity.


\begin{figure}[!h]
\begin{center}
\includegraphics[trim = 30mm 0mm 0mm 0mm, width=8cm]{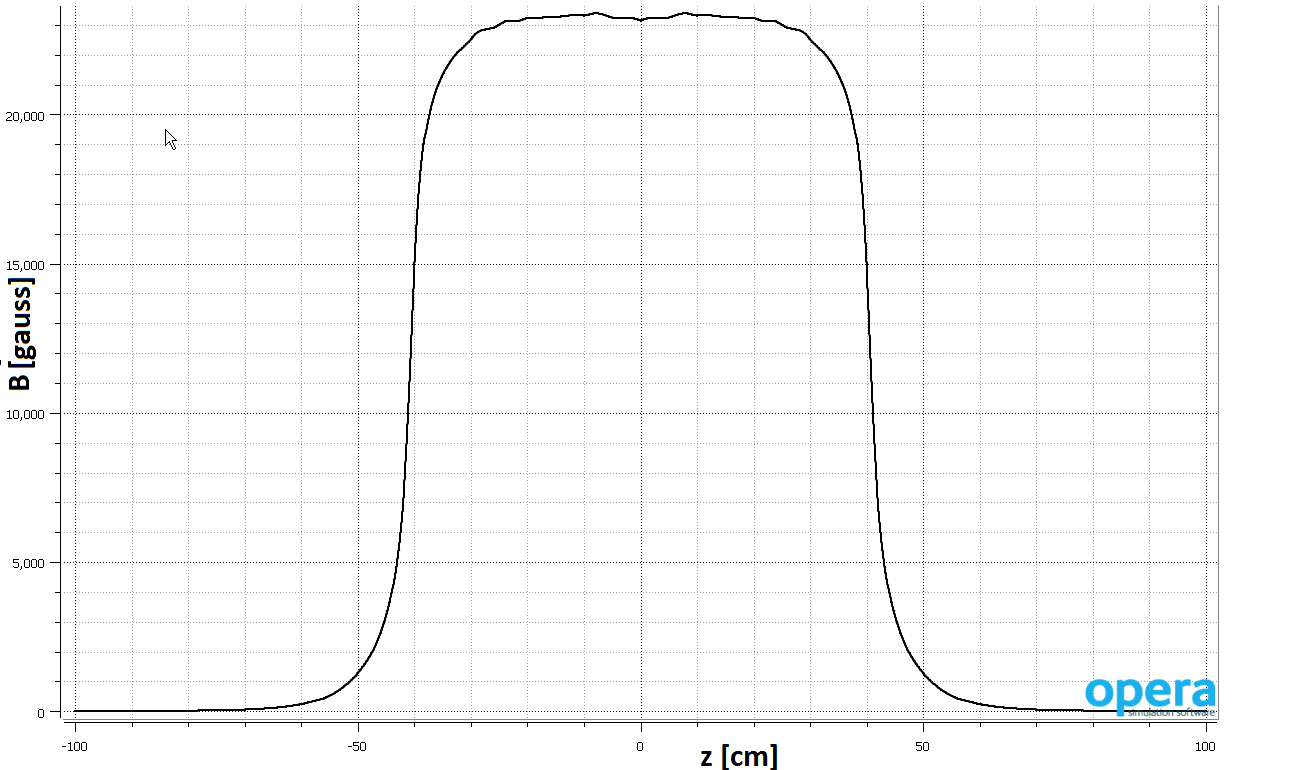}
\end{center}
\caption{The vertical field in the magnet along the beam line.} 
\label{fig:field}
\end{figure}

\begin{figure}[!h]
\begin{center}
\includegraphics[trim = 0mm 0mm 0mm 0mm, width=14cm]{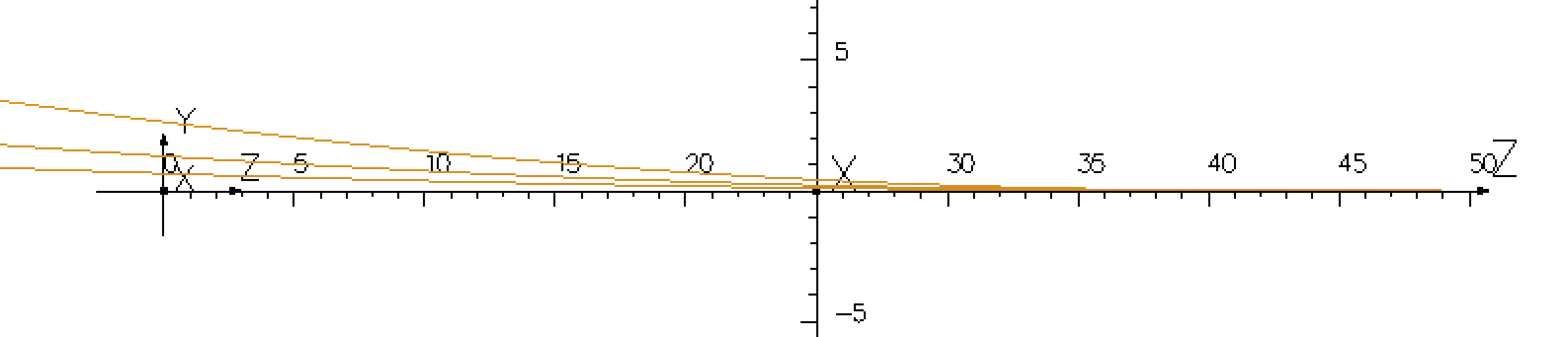}
\end{center}
\caption{The trajectories of electrons with momenta 8.8, 4.4, and 2.2 GeV.} 
\label{fig:tracks-a}
\end{figure}

\begin{figure}[!h]
\begin{center}
\includegraphics[trim = 10mm 10mm 10mm 10mm, width=14cm]{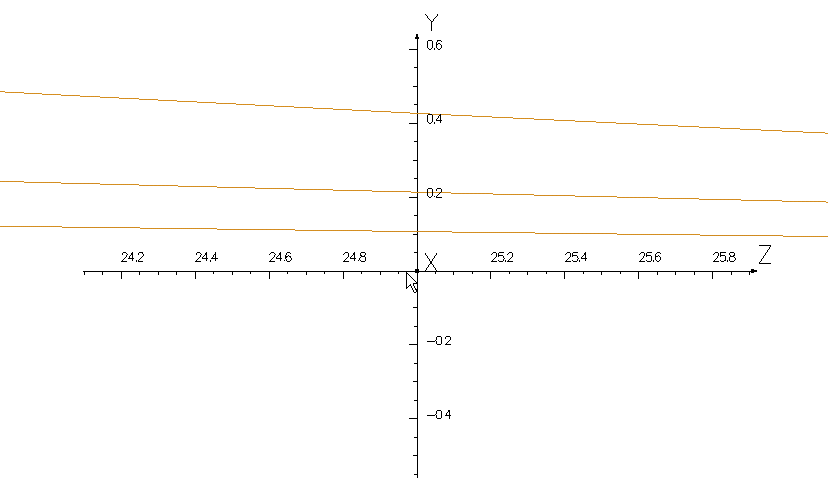}
\end{center}
\caption{Zoomed version of the previous figure.} 
\label{fig:tracks-b}
\end{figure}

The transverse field along the beam direction on the magnet axis is shown in Fig.~\ref{fig:field}.
The resulting trajectories for 8.8, 4.4, and 2.2 GeV electrons are shown in Figs.~\ref{fig:tracks-a} and
\ref{fig:tracks-b}.

\clearpage

\subsection{Diffuser-Absorber}

\begin{figure}[htbp]
\begin{minipage}[h]{0.55\textwidth}
\begin{center}
\includegraphics[trim = 0mm 0mm 0mm 0mm, width = 0.75\textwidth]{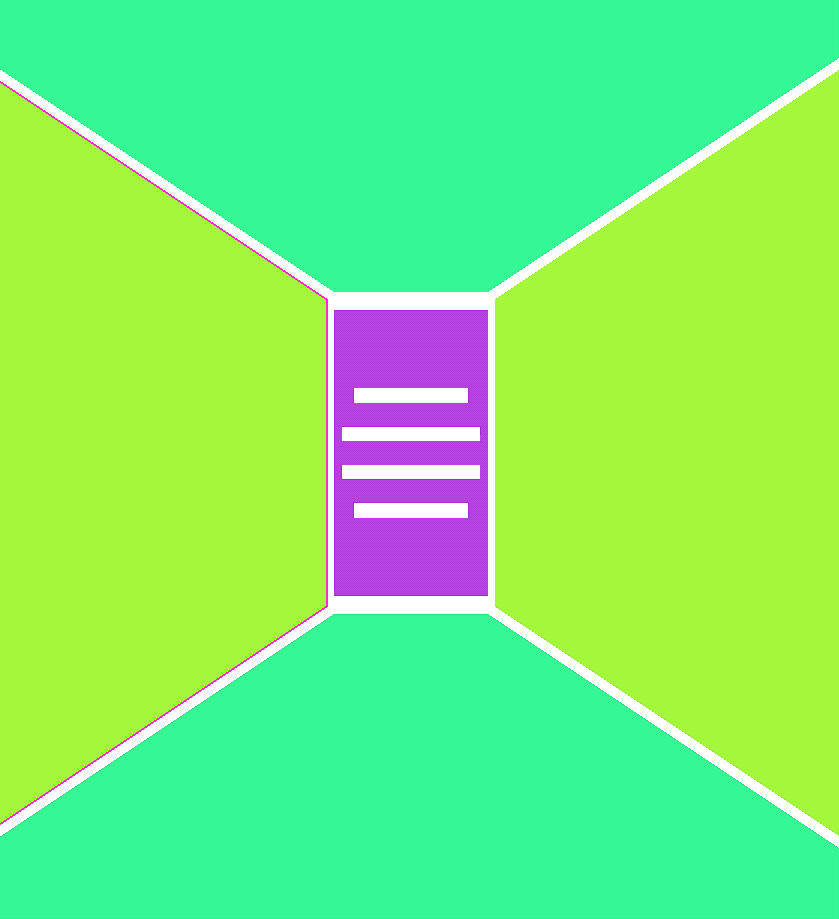}
\end{center}
\end{minipage}
\hskip .25 in
\begin{minipage}[h]{0.35\textwidth}
\begin{center}
\caption{Front view of the magnet gap filled with copper blocks and 
the CuW80 insert with horizontal slots. All four slots for the beam are shown.
The magnetic field is horizontal.} 
\label{fig:insertS}
\end{center}
\end{minipage}
\end{figure}

\ind There are two options for the diffuser-absorber.
The first one is a rotating cylinder with a set of 2-mm diameter holes. 
This is the best for beam collimation but complicates the cooling water arrangement.
The second is a vertically oscillating block with horizontal slots 
for the beam as shown in Fig.~\ref{fig:insertS}.
The MC result shows no increase in radiation in the second option compared with the first one
in spite of a wider channel, so we will discuss below only the second option.

The large length of the insert (1.25~meters) allows almost complete blocking of the radiation
created inside the insert mostly at a distance of 40-60 cm from the entrance.
The radiator will be mounted at a distance of 5-10 cm from the entrance inside the insert.
The oscillating option has the advantage of design simplicity, especially for the cooling system.
At the same time, the slots will help to arrange uniform distribution of the heat 
load in the diffuser-absorber.

Per our calculation, of the 10~kW total power of the beam, about 8 kW will be
deposited in the diffuser-absorber and the rest in the copper wedges between 
the magnet poles, see Fig.~\ref{fig:insertS}.
The absorber-diffuser will be made from a heavy metal alloy (CuW80) whose
short radiation length and high density will reduce the size of the radiation source.

A water flow will be used to cool off the inserts with a closed loop chiller.
The water line will use stainless steel bellows which are capable of holding $10^9$ cycles of deformation.
A 3-4 m long metal shaft will be used to connect the absorber-diffuser to a step-motor 
to activate the motion.

\vskip 0.5 in

\section{Radiation issue and its mitigation}
\label{sec:radiation}

The radiation aspects of the proposed device could be fully analyzed via
a GEANT4-based MC simulation as it is presented in Section~\ref{sec:MC_results}.
However, the analysis presented in this section helps to clarify 
the formulation of the concept and supports the validation of the MC results.

\subsection{Radiation from a thick target}

Radiation generation by a high energy electron beam has been investigated since the 1960s
after the construction of the SLAC and DESY accelerators.
The variation of radiation intensity per kW of beam power (mostly neutron) is very small
for beam energies above 500 MeV.
The PDG review~\cite{PDG} shows a plot with energies up to 100 MeV, and
the original SLAC report~\cite{Swanson} discussed 500 and 1000 MeV beam energies.

\begin{figure}[htbp]
\begin{minipage}[h]{0.55\textwidth}
\begin{center}
\includegraphics[width= 10 cm]{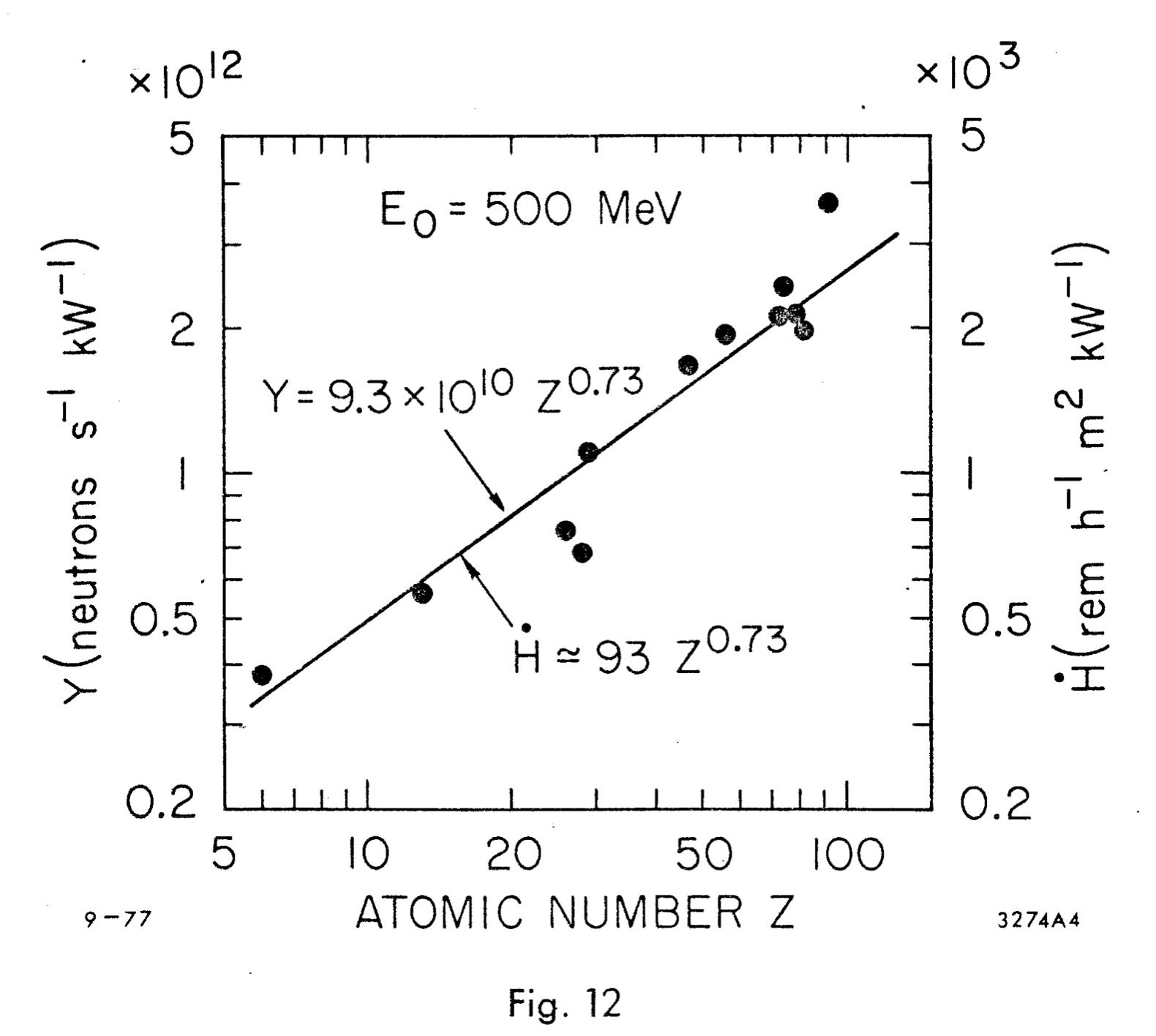} 
\end{center}
\end{minipage}
\hskip 0.5 in
\begin{minipage}[h]{0.35\textwidth}
\begin{center}
\caption{Neutron yield from a thick target induced by 
a high energy electron beam (from the SLAC report~\cite{Swanson}).}
\label{fig:SwansonFig}
\end{center}
\end{minipage}
\end{figure}

\begin{figure}[htbp]
\begin{minipage}[h]{0.55\textwidth}
\begin{center}
\includegraphics[trim = 0mm 0mm 0mm 0mm, width = 0.75\textwidth]{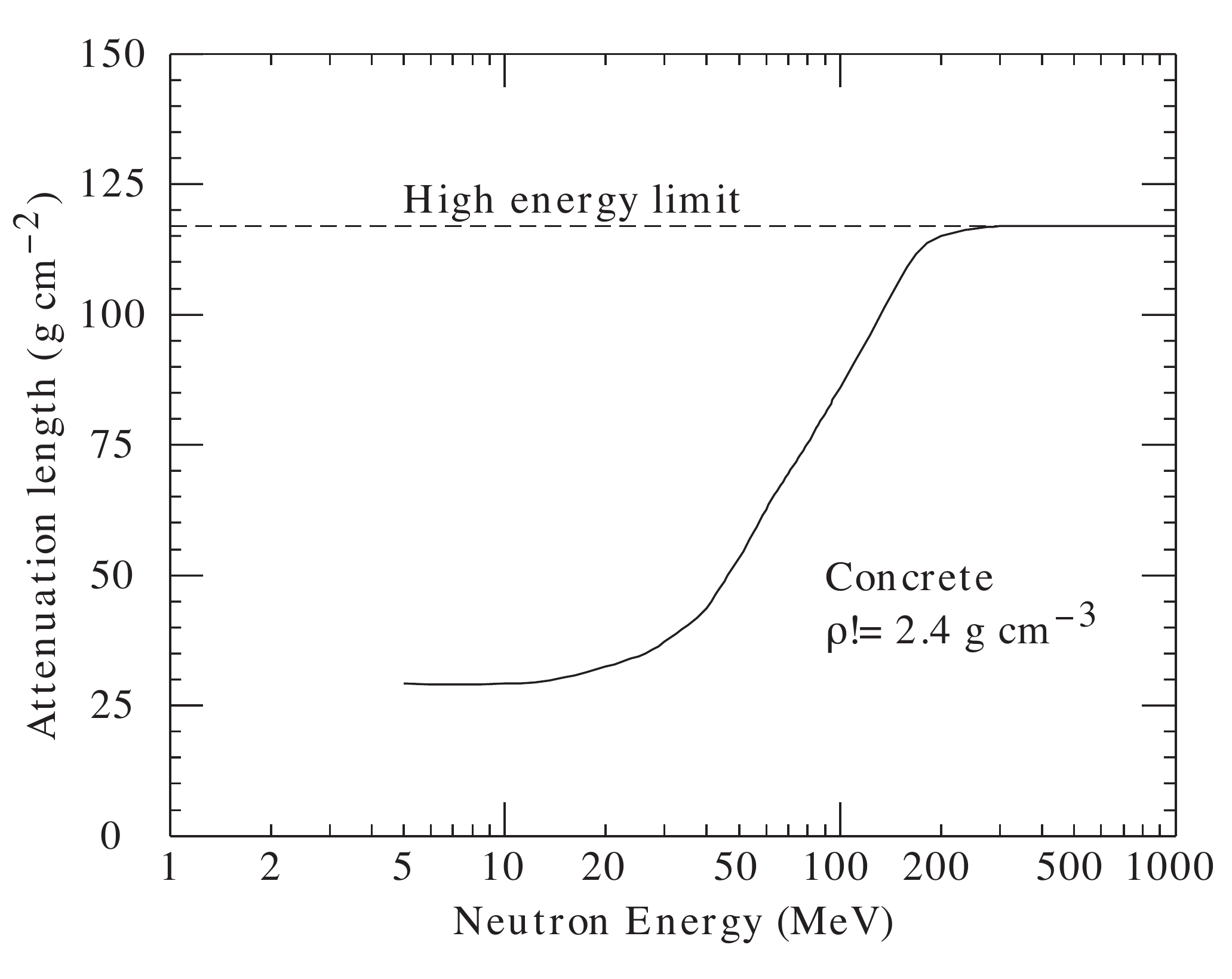} 
\end{center}
\end{minipage}
\hskip 0.5 in
\begin{minipage}[h]{0.35\textwidth}
\begin{center}
\caption{Attenuation length of the neutron flux vs. neutron energy (from the PDG report~\cite{PDG}).}
\label{fig:PDG_lambda}
\end{center}
\end{minipage}
\end{figure}

Figure~\ref{fig:SwansonFig} shows the neutron yield and related radiation dose 
(at a one-meter distance from the target).
In this project we will use a beam of 1.2~$\mu$A at 8.8~GeV energy with a total power of 10.5~kW.
An unshielded target of large thickness irradiated with such a beam will release 
a total flux of $2\times10^{13}$ neutrons per second, which leads
to a radiation level of 20~kRad/hour at a distance of one meter from the target, as one can
see from the right scale in the same Figure.
It is also easy to find that the iron yoke of the magnet and external shielding should be
capable of reducing this radiation level by 3-4 orders of magnitude.
Indeed, as it is shown in Fig.~\ref{fig:PDG_lambda}, taken from the PDG report,
a concrete slab of 11~cm thickness reduces the neutron flux by a factor of $e$.
A reduction factor of 1000 is achievable with a concrete slab of 75~cm.
High energy neutrons with energies above 200~MeV have a 4 times longer
attenuation length up to 120 g/cm$^2$.
However, even for such neutrons, with the proposed shielding thickness of 100 cm 
of iron, the reduction is expected to be by a factor of 700.

\begin{figure}[htbp]
\begin{minipage}[h]{0.75\textwidth}
\begin{center}
\includegraphics[trim = 0mm 0mm 0mm 0mm, width = 1.\textwidth]{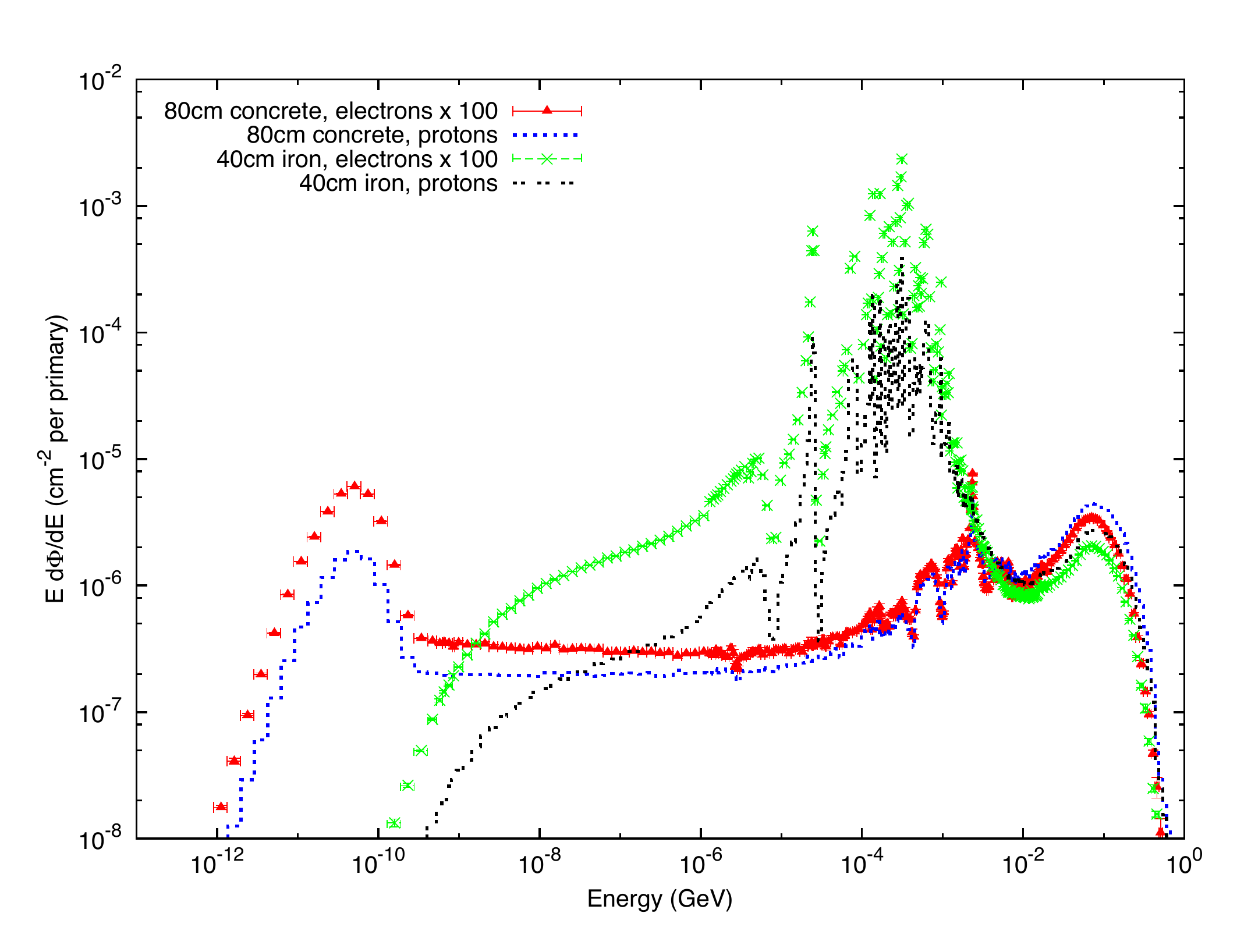} 
\end{center}
\end{minipage}
\hskip 0.15 in
\begin{minipage}[h]{0.2\textwidth}
\begin{center}
\caption{The spectrum of neutrons from a thick copper target
bombarded by a 25~GeV electron beam (from the PDG report~\cite{PDG}).
The rates are outside the 80~cm concrete shielding or 40~cm iron shielding
at 90 degrees from the beam direction.}
\label{fig:PDG_fluka}
\end{center}
\end{minipage}
\end{figure}

Additional insight into the shielding of radiation and the neutron rate outside the shielding could be
obtained from Fig.~\ref{fig:PDG_fluka}, which shows the spectra of neutrons
outside of the shielding for a 25 GeV electron beam striking a thick copper target.
A combination of iron and concrete shielding should provide the best result.
The neutron intensity integrated over the energy spectrum is consistent with
the information in Fig.~\ref{fig:SwansonFig}.

\subsection{Hermeticity of shielding}
\ind In the design of the shielding, some attention should also be on the hermeticity of the 
shielding structure.
For example, the penetration of a 15-cm diameter is acceptable for a 100-cm long  
tunnel in a 90 degree direction. 
Diffusion of the neutrons through such a channel is strongly suppressed
because of the large number of the neutron reflections required (45 for such a tunnel) 
for the diffusive transport and the low albedo factor of the reflection (typically between 0.3 and 0.5).
The solid angle of such a large opening is 0.0016 of the total 4$\pi$,
so the channel should not originate in the hottest spot of the diffuser-absorber
and should use a chicane if possible. 
However, in the forward direction (along the beam), the opening in the shielding needs 
to be much smaller than the discussed 15-cm diameter channel. 
It must be minimized because the high energy neutron 
distribution has a small opening angle of $2 m_\pi/E_\gamma$.
For the proposed photon beam channel (2 mm x 20 mm), the solid angle is of $3 \times10^{-5}$,
which is sufficiently small to block 98\% of the most energetic neutrons. 

Using the considerations above, we find that at a typical distance of 15~meters
in Hall A between the source and the sensitive electronics, the neutron dose
would be on the order of 100 mRad per hour, which is acceptable according to
previously performed experiments in Hall A with an open geometry detector package.

In spite of the optimistic results from the estimates and considerations shown above, 
already in the proposal stage we decided to develop a full MC simulation of the radiation 
using a MC code in the GEANT4 framework, which is described in the next sections.

\section{Monte Carlo simulation}
\label{sec:mc}
\ind  While the simplified beam deflection calculation shown above indicates that the design is capable of 
deflecting the primary electron beam into the shielding part of the device, substantially more insight can be 
gained from a detailed Monte Carlo simulation. 
In order to gauge the expected performance of the proposed setup and to provide a solid basis for 
the expected dose rate calculation needed to ensure safety, the whole sweeper magnet/target dump 
(and its shielding) were implemented in a GEANT4--based ({\bf g4}) simulation and extensively tested.

\subsection{Monte Carlo code configuration}
\label{sec:mc_general}

A list of the code design choices made in implementing the sweeper magnet/target dump model into 
the simulation are listed (and justified, where appropriate) below:

\begin{itemize}
\item {{\bf Version:} Geant4, v. 10.1 (release date Dec. 2014) implemented on CentOS Linux.}
\item {{\bf Multithread:} Given the relative simplicity of the model, 
multithreading, though available, was not used.}
\item {{\bf Magnetic Field Model:} The magnetic field map (with a 2 cm mesh size) produced by TOSCA was read at the beginning of the program and used in the simulation. A 3D linear interpolation function was implemented and used in the code. 
For convenience, a ``field strength" variable stored in a separate text file was read at runtime, allowing for the scaling 
of the magnetic field without recompilation/extensive changes in the field map. 
All results quoted here were obtained using the ``nominal" value of the magnetic field.}
\item {{\bf Geometry:} In order to keep the complete definition of the models' geometry (and materials) 
completely separated from the source code of the simulation, 
the GDML~\cite{gdml} extension of GEANT4 was used. 
As all of the geometry is contained in a stand-alone xml file read at runtime, 
the overhead and potential pitfalls (usually) associated with geometry and/or material changes 
in the simulation were greatly diminished.}
\item {{\bf Materials:} The code makes extensive use of the NIST--defined materials already defined in g4. In a few cases, materials were defined in the gdml file. 
For example, the JLab--developed boron--rich concrete was considered for use as a shielding material 
(as a cost--saving measure, though we decided not to use it).}
\item {{\bf Random Number Generator:} CLHEP's ``RanEcuEngine"~\cite{ranecu}.
Starting from the same pair of seeds, this random number generator has a period of $10^{18}$, 
which was deemed sufficient for this simulation. 
Furthermore, the pairs of random seeds were changed after every $10^5$ generated beam electrons.}
\item {{\bf Physics List:} 
The ``physics list" (i.e. a list of all the particle types and processes that they might undergo, for all energy scales, that are enabled in g4) used was {\bf FTFP--BERT}. 
This model combines the {\it Bertini Cascade Model}~\cite{bertini, g4paper}, known to reproduce detailed cross--section data for nucleons as well as pions and kaons in the $\sim$ GeV range 
(nominally 0 to 5~GeV) with the {\bf Fritiof Model}~\cite{fritiof, g4paper}, expected to be 
valid for larger ($>$5~GeV) energies. }
\item {{\bf Input/Output:} Besides the gdml geometry file and the field map mentioned above, the code also 
reads a plain text file that defines the energy, location, direction, and spread of the initial electron beam. 
The output consists of diagnostic messages, plain text reports (dose rates, particle flux through various parts of the model, etc.) and a ROOT tree containing the initial ``gun" particle (8.8 GeV electrons for proposal PR12-15-003) as well as a list of all particles (four momenta, position, PDG ID, etc.) 
exiting the magnet area, for further analysis.}
\item {{\bf Scoring volumes:} For dose rate calculations, the vacuum ``ghost" volumes placed 
15~m from the center of the magnet were defined and monitored. 
Additional volumes were used at the external surface of the shielding.}
\end{itemize}

\subsection{Event generator validation}
\label{sec:validation}

\begin{figure}[tbp]
\begin{center}
\includegraphics[trim = 0mm 0mm 0mm 0mm, width = 0.75\textwidth]{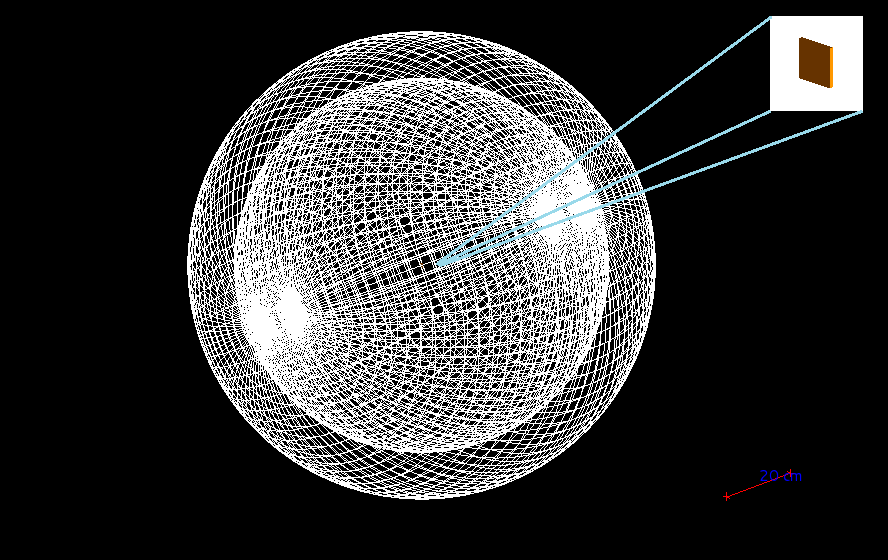} 
\caption{The GEANT model for calculation of the particle yields from a 1-mm thick $^{12}$C target.}
\label{fig:C_source}
\end{center}
\end{figure}

In order to gain confidence that, as implemented, the Monte Carlo simulation of the magnet generates particles 
at rates that are similar to accepted standards in the field, the following simple exercise was carried out:
\begin{itemize}
\item {All the geometry of the magnet and its shielding was stripped away. 
The use of gdml greatly simplified this process.}
\item {A 1-mm Carbon target was placed in the center of the g4 ``world" 
with a spherical bounding envelope around it, as seen in Figure~\ref{fig:C_source}.}
\item {To match the existing~\cite{Pavel} particle production cross--section, 
the initial electron beam energy was set to 12~GeV.}
\item {A substantial number ($1.2 \times 10^9$) of electrons were fired at the carbon target. 
Emerging particles were saved in ROOT trees as explained earlier.}
\item {Particle production rates as a function of the kinetic energy and angle (normalized to the number of electrons simulated) were obtained. 
Results for $e^- + C \to n^- + X$ are shown in Figure~\ref{fig:Gabriel_C_n}.} 
\item {The particle rates we obtained from our GEANT4-generated events compare 
well with \lq\lq{}standard\rq\rq{} particle production rates~\cite{Pavel} in both 
the magnitude and shape. 
This gives us confidence that the Monte Carlo simulation does not underestimate production rates
and, therefore, that the dose rate calculation based on this model should be a reasonable estimate.}
\end{itemize}

\ind  The key step in the validation of the code is the test of the particle production rate,
which is controlled by a number of switches and often could be incorrect.
Below we compare the results of our code with the well-tested DINREG results.

\subsubsection{MC calculation of the particle production in the DINREG code}

\ind The DINREG code in the Geant-3 framework was developed by P.~Degtiarenko
for the calculation of particle yields and radiation doses in the experimental halls at JLab.
The results of this code were compared with numerous measurements at JLab
and published data.
A systematic difference was observed at JLab: DINREG overpredicts the radiation
level measured in the hall by a factor of 3.
The origin of this factor is not completely known but could be due to the calibration
method used for the radiation monitors.
\begin{figure}[tbp]
\begin{center}
\includegraphics[trim = 20mm 20mm 50mm 10mm, width = 0.72\textwidth]{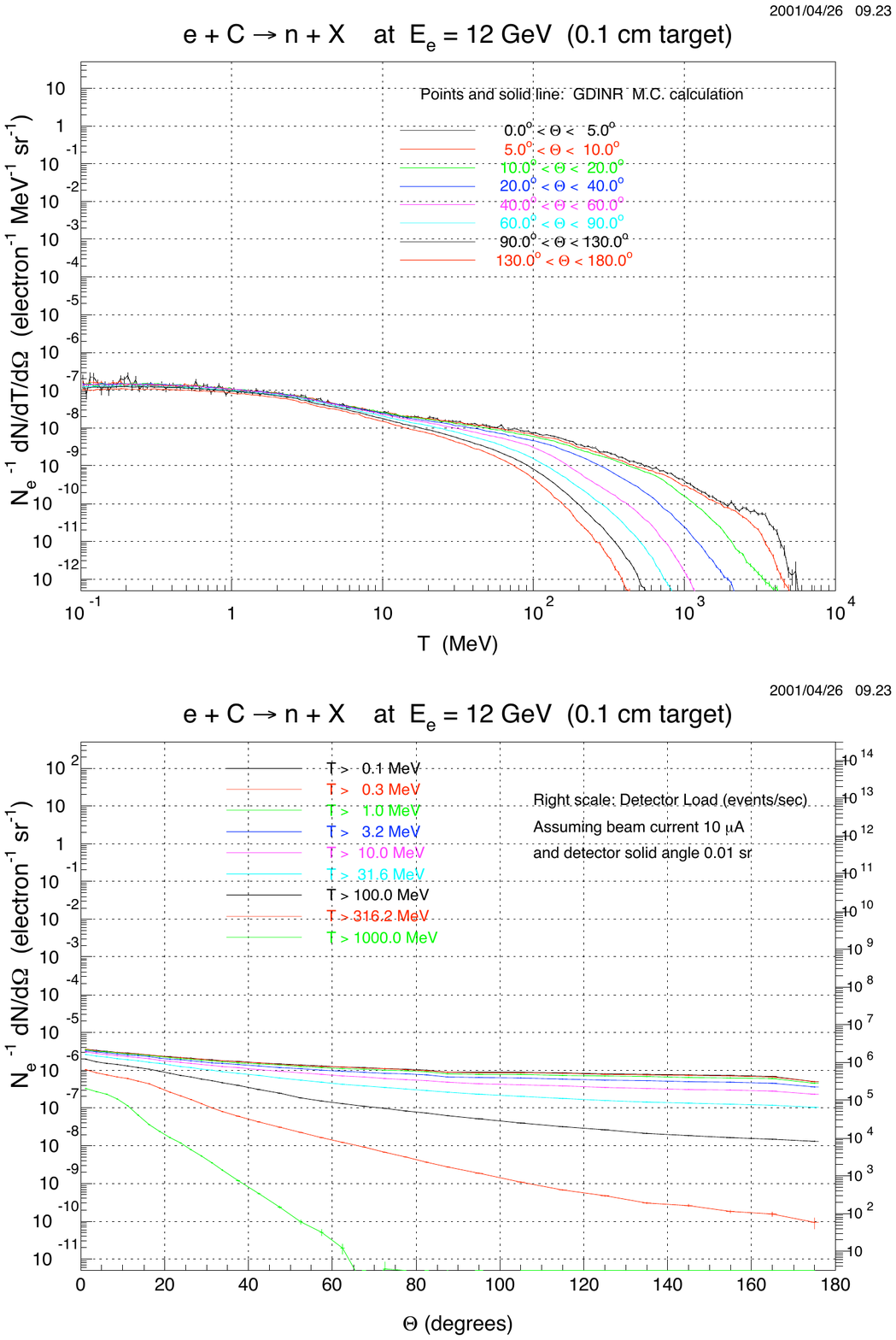} 
\caption{The intensity of the neutrons produced in a 1-mm Carbon target by 
a 12-GeV electron beam~\cite{Pavel}.}
\label{fig:DINREG_C_n}
\end{center}
\end{figure}

\subsubsection{MC calculation of the particle production in the GEANT4 code}

\ind Figure~\ref{fig:DINREG_C_n} shows the neutron yield for a 1-mm Carbon target
with a 12-GeV electron beam.
This source yield was also found to be in good agreement with FLUKA calculations.
Because DINREG uses the GEANT package for the geometry and the particle
interactions and the tracing of particles through the material, 
it is very much suitable for experiment simulations.
However, the currently more common package, GEANT4, was used for
the MC simulation of the proposed device.
We are using the result shown in Fig.~\ref{fig:DINREG_C_n} 
to validate our MC code based on the GEANT4 framework.
\begin{figure}[!h]
\begin{center}
\includegraphics[trim = 0mm 0mm 0mm 0mm, width = 0.85\textwidth]{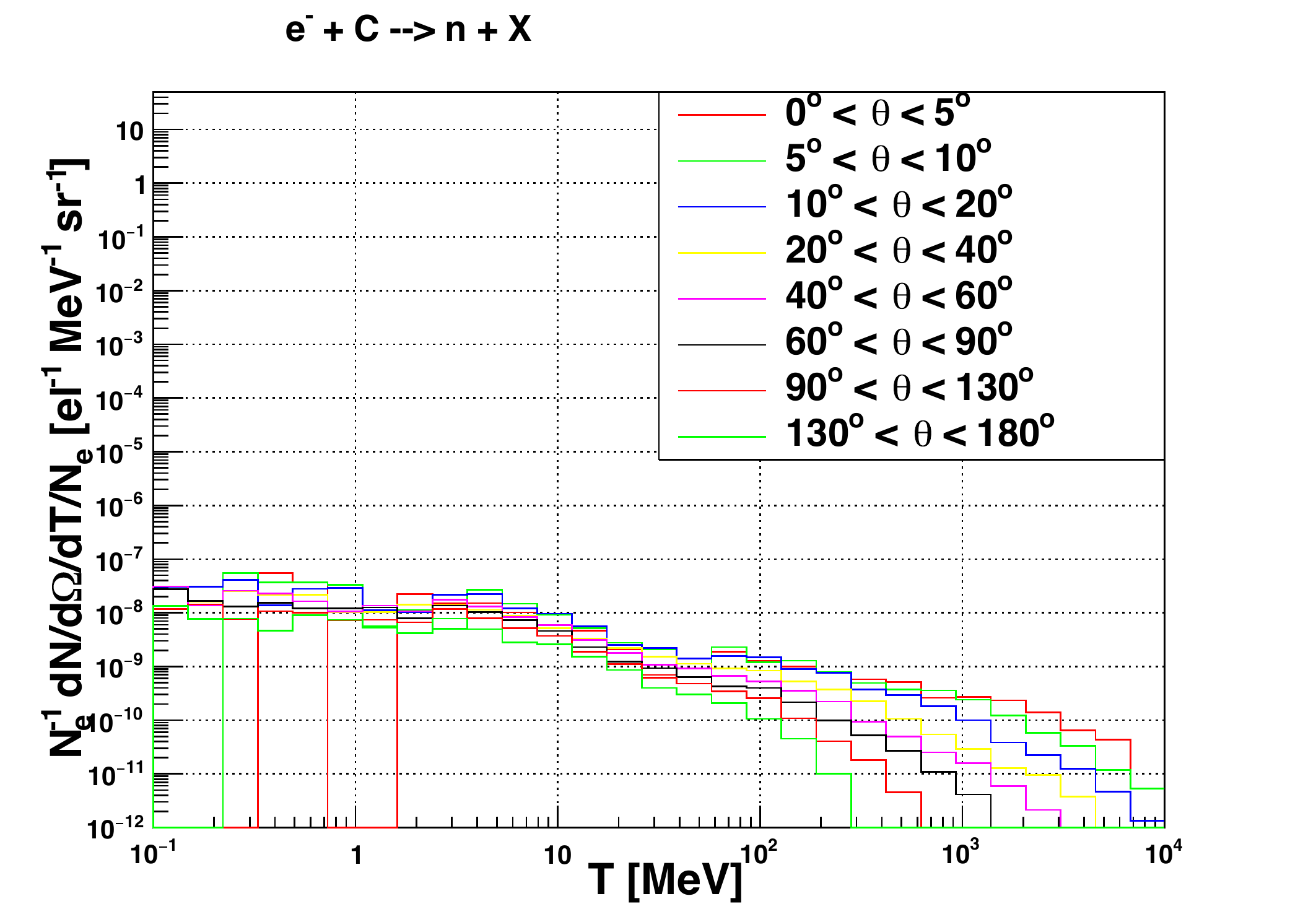} 
\includegraphics[trim = 0mm 0mm 0mm 0mm, width = 0.85\textwidth]{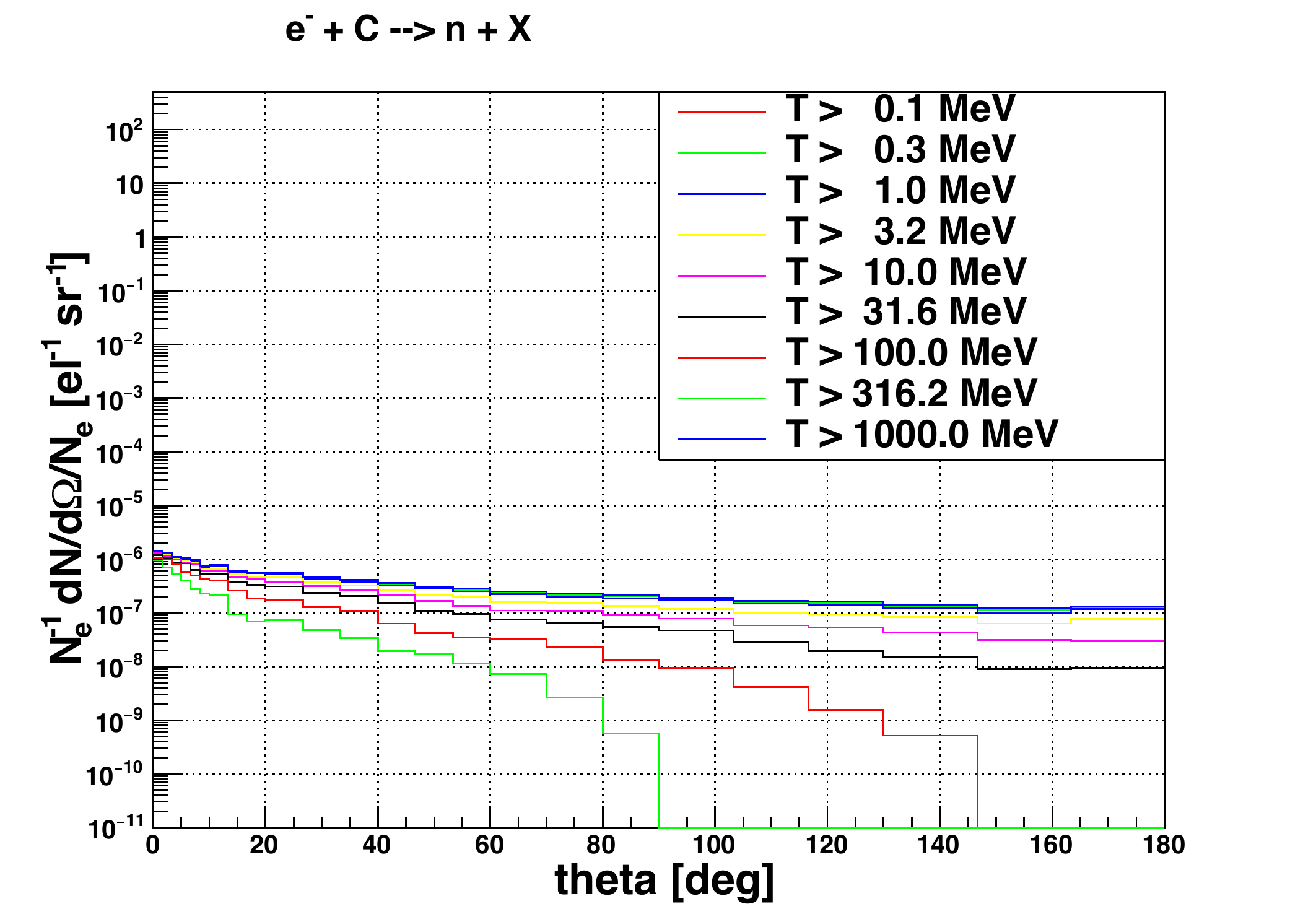} 
\caption{The intensity of the neutrons produced in a 1-mm Carbon target
by a 12-GeV electron beam from our GEANT4-based code.}
\label{fig:Gabriel_C_n}
\end{center}
\end{figure}

The MC was performed using the MC code developed for the Compact Photon
Source simulation but with all the device geometry replaced by a 1-mm Carbon plate.
The resulting rates are shown in Fig.~\ref{fig:Gabriel_C_n}.
The spectra are very similar to DINREG's except for very large neutron energy 
where the DINREG statistical approach to the particle energy is unlikely to be applicable.
In the angular range below 45 degrees and the energy up to a few hundreds of MeV, 
our MC code predicts a lower flux of neutrons by a factor of 2 than DINREG does.
Such agreement is sufficiently good for the purpose of our preliminary study.


\subsection{Geometry of the magnet-dump model in MC}

The GEANT4 model of the Compact Photon Source is shown in the set of pictures below.
\begin{figure}[!htbp]
\begin{minipage}[h]{0.65\textwidth}
\begin{center}
\includegraphics[trim = 0mm 0mm 0mm 0mm, width = 0.65\textwidth]{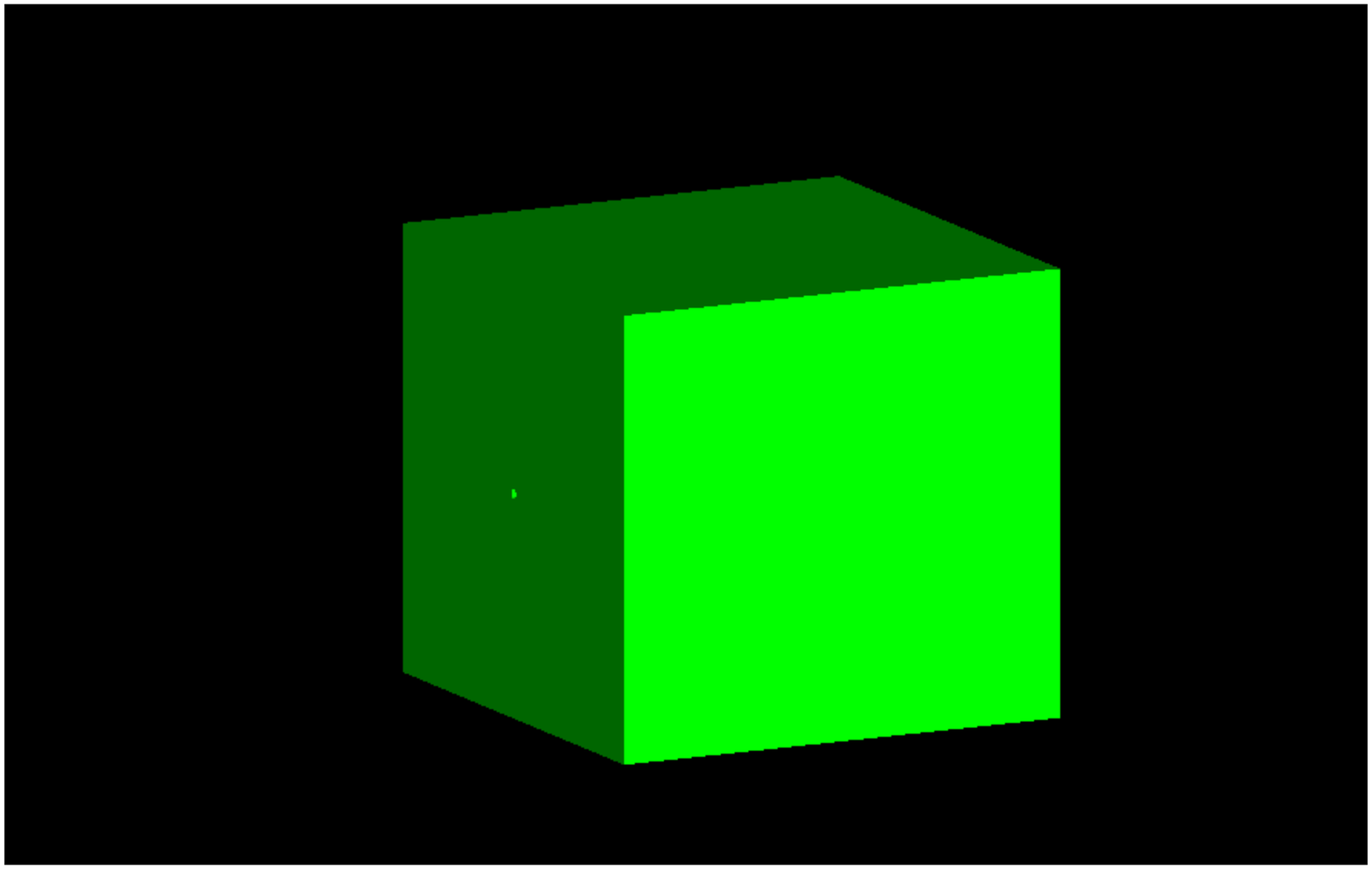} 
\end{center}
\end{minipage}
\hskip 0.01 in
\begin{minipage}[h]{0.35\textwidth}
\begin{center}
\caption{The external view of the shielding (the overall size is about 2.1~m x 2.2~m x 2.6~m).}
\label{fig:PPS_view_1}
\end{center}
\end{minipage}
\end{figure}
\begin{figure}[!htbp]
\begin{minipage}[h]{0.65\textwidth}
\begin{center}
\includegraphics[trim = 0mm 0mm 0mm 0mm, width = 0.65\textwidth]{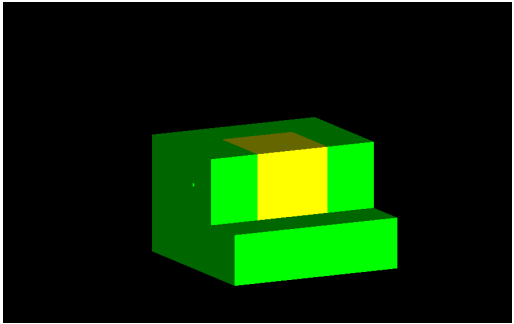} 
\end{center}
\end{minipage}
\hskip 0.01 in
\begin{minipage}[h]{0.35\textwidth}
\begin{center}
\caption{The view of the partially open shielding. 
The concrete blocks around the magnet are shown in yellow. 
The downstream shielding has a 0.66 m thickness.}
\label{fig:PPS_view_2}
\end{center}
\end{minipage}
\end{figure}
\begin{figure}[!htbp]
\begin{minipage}[h]{0.65\textwidth}
\begin{center}
\includegraphics[trim = 0mm 0mm 0mm 0mm, width = 0.65\textwidth]{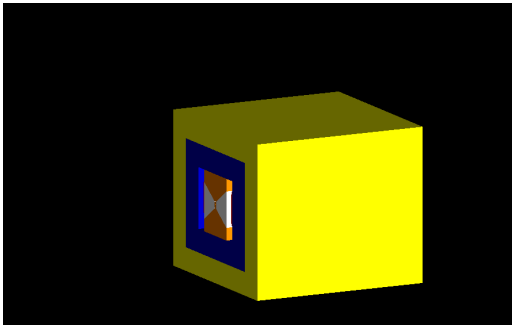} 
\end{center}
\end{minipage}
\hskip 0.01 in
\begin{minipage}[h]{0.35\textwidth}
\begin{center}
\caption{The view of the magnet yoke (shown in blue) and the concrete shielding (shown in yellow). 
Copper blocks fill the space inside the magnet.}
\label{fig:PPS_view_3}
\end{center}
\end{minipage}
\end{figure}
\begin{figure}[!htbp]
\begin{minipage}[h]{0.65\textwidth}
\begin{center}
\includegraphics[trim = 0mm 0mm 0mm 0mm, width = 0.65\textwidth]{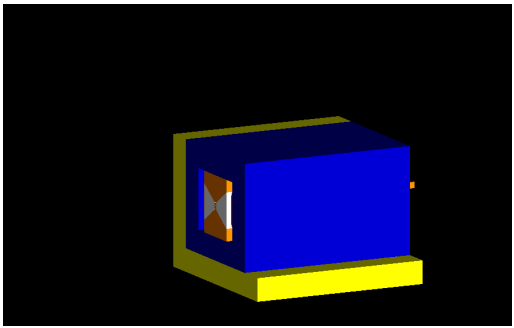} 
\end{center}
\end{minipage}
\hskip 0.01 in
\begin{minipage}[h]{0.35\textwidth}
\begin{center}
\caption{The view of the magnet with the top and side concrete plates removed.}
\label{fig:PPS_view_4}
\end{center}
\end{minipage}
\end{figure}
\begin{figure}[!htbp]
\begin{minipage}[h]{0.65\textwidth}
\begin{center}
\includegraphics[trim = 0mm 0mm 0mm 0mm, width = 0.65\textwidth]{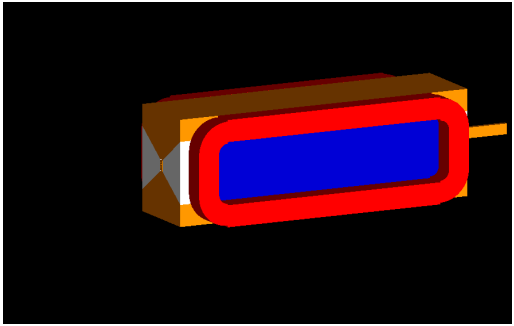} 
\end{center}
\end{minipage}
\hskip 0.01 in
\begin{minipage}[h]{0.35\textwidth}
\begin{center}
\caption{The view of the coils and the poles (shown in blue). 
The grey blocks are the copper inserts with the same side profile as the magnet poles.}
\label{fig:PPS_view_6}
\end{center}
\end{minipage}
\end{figure}
\begin{figure}[!htbp]
\begin{minipage}[h]{0.65\textwidth}
\begin{center}
\includegraphics[trim = 0mm 0mm 0mm 0mm, width = 0.65\textwidth]{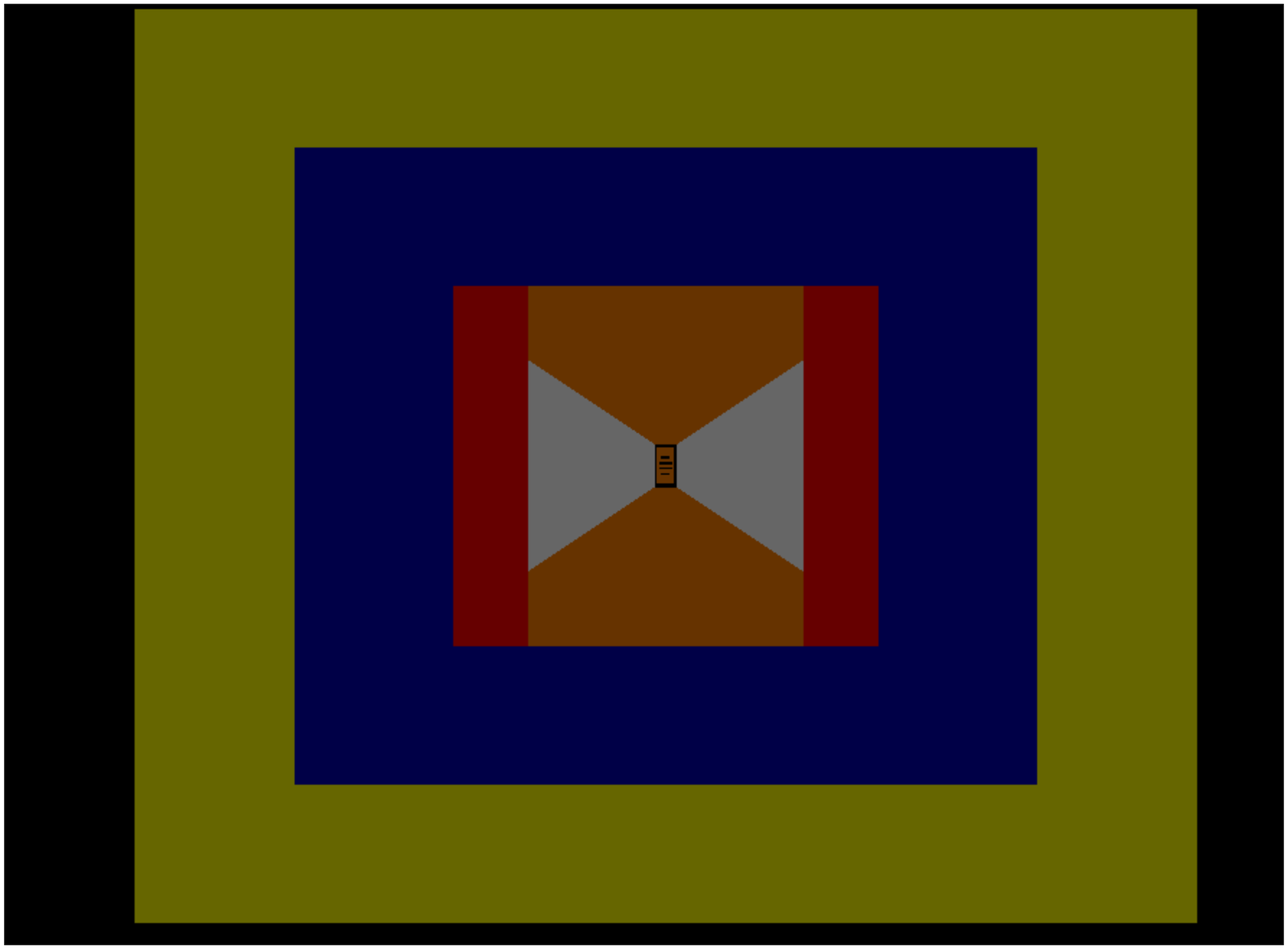} 
\end{center}
\end{minipage}
\hskip 0.01 in
\begin{minipage}[h]{0.35\textwidth}
\begin{center}
\caption{The front view of the magnet and the concrete shielding.
The insert between the poles is the diffuser-absorber with 
the beam rastering area shown in dark red.}
\label{fig:PPS_view_7}
\end{center}
\end{minipage}
\end{figure}

\section{Radiation calculation results}
\label{sec:MC_results}

\subsection{The dose rates at key locations}

This section presents the counting rates, energy spectra, and induced radiation dose rates 
for the neutrons and gamma-rays in several key locations:
\begin{itemize}
\item The coil of the magnet-dump.
\item The coil of the polarized target magnet.
\item 15 meters from the source.
\item The polarized target cell.
\end{itemize}

As shown in the previous section, the event generator used in the Monte Carlo 
simulation of the sweeper magnet-beam dump combination matches particle production 
rates accepted by the community for the energy range of interest. 
One can then proceed to simulate the proposed magnet geometry to assess the suitability of the design 
for providing the clean, narrow photon beam the experimental program calls for and also to determine 
the size and materials that need to be used as shielding around the magnet in order to bring down 
the radiation levels to acceptable JLab and DOE levels.

\subsubsection {Radiation at the magnet-dump coils}

\begin{figure}[!htbp]
\begin{center}
\includegraphics[trim = 0mm 0mm 0mm 0mm, width = 0.7\textwidth]{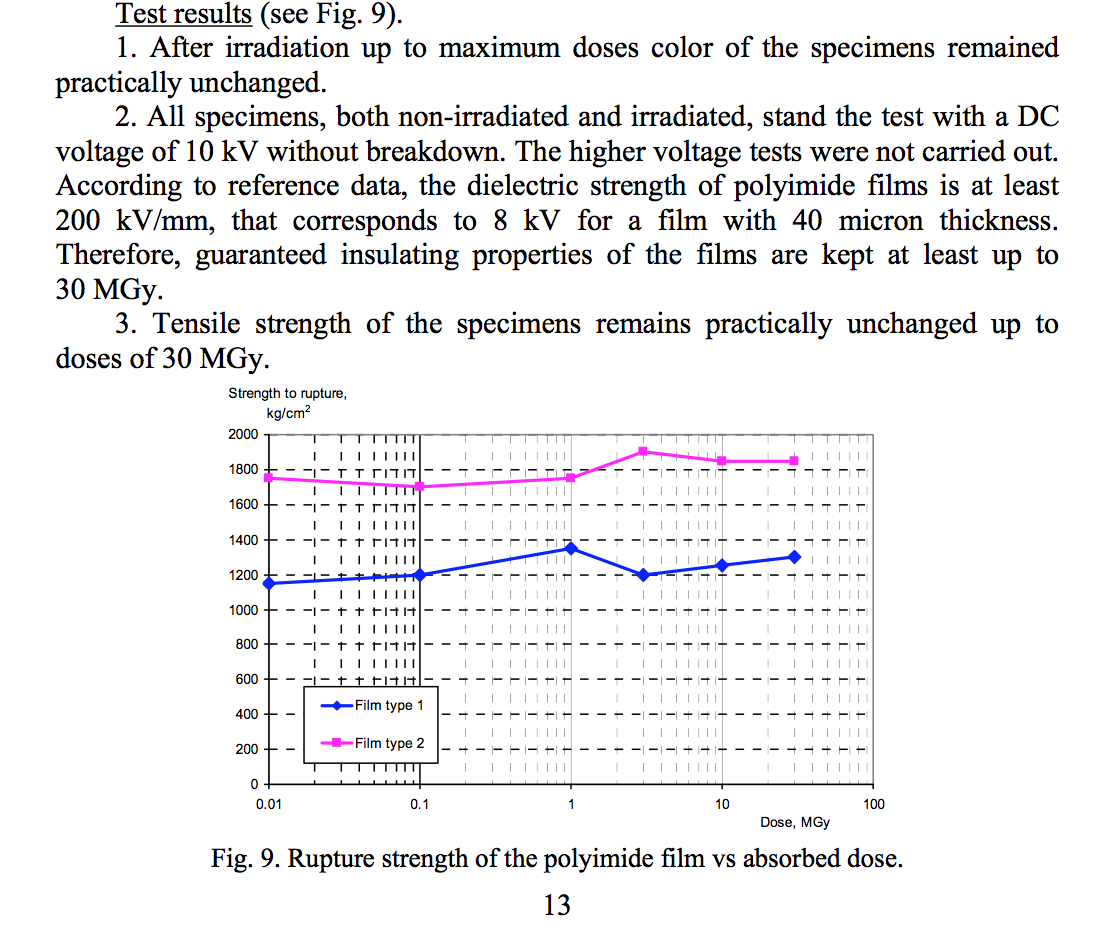} 
\end{center}
\caption{The test results of the kapton tape properties vs radiation dose~\cite{Kapton}.}
\label{fig:kapton}
\end{figure}

The dose rate on the coils was found to be 80~kRad/hour, which leads 
to a total dose of 20~MRad over the duration of the WACS experiment. 
Such a high dose is not unusual for an extraction line in proton accelerators.
For the proposed device, it would be sufficient to replace the epoxy insulator of 
the coil winding used in a low radiation environment with a kapton tape.
The results of the kapton film investigation are shown in Fig.~\ref{fig:kapton}.

\subsubsection {Radiation at a source shielding surface}

The neutron induced dose rate was found to be 1.2 rem/hour at the top, 1.3 rem/hour at the bottom,
2.1 rem/hour on the left, 2.7 rem/hour on the right, and 3.5 rem/hour at the front of the shielding.
Such a level of radiation should not lead to significant residual radiation after the beam is 
switched off.

\subsubsection {Radiation at a 15-meter distance}

The neutron induced dose was evaluated by means of the large detectors located 15 meters from the radiator.
We observed average doses of 15-20 mrem/hour in 90-degree directions,
about 90 mrem/hour in the backward direction, and 270 mrem/hour in the forward direction.
In the forward direction the photon distribution is strongly peaked at the beam line as shown below, 
so it is discussed in detail later.

\begin{figure}[!htbp]
\begin{center}
\includegraphics[trim = 0mm 0mm 0mm 0mm, width = 0.8\textwidth]
{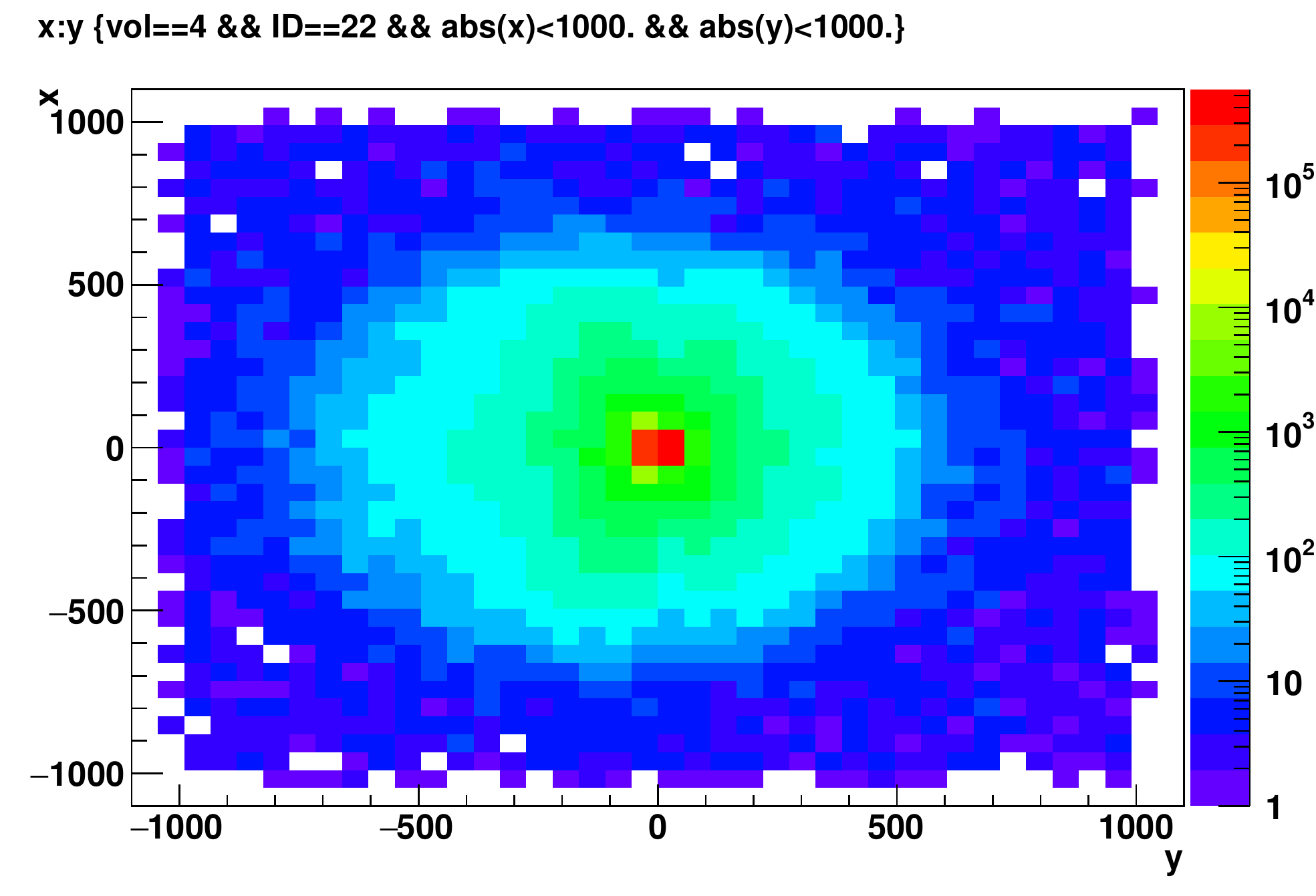} 
\end{center}
\caption{The first zoomed hit distribution in the forward direction at a distance of 15 m from the radiator
for photon energies above 1 keV. The coordinates X and Y are in millimeters.}
\label{fig:hits15b}
\end{figure}

\begin{figure}[!htbp]
\begin{center}
\includegraphics[trim = 0mm 0mm 0mm 0mm, width = 0.8\textwidth]
{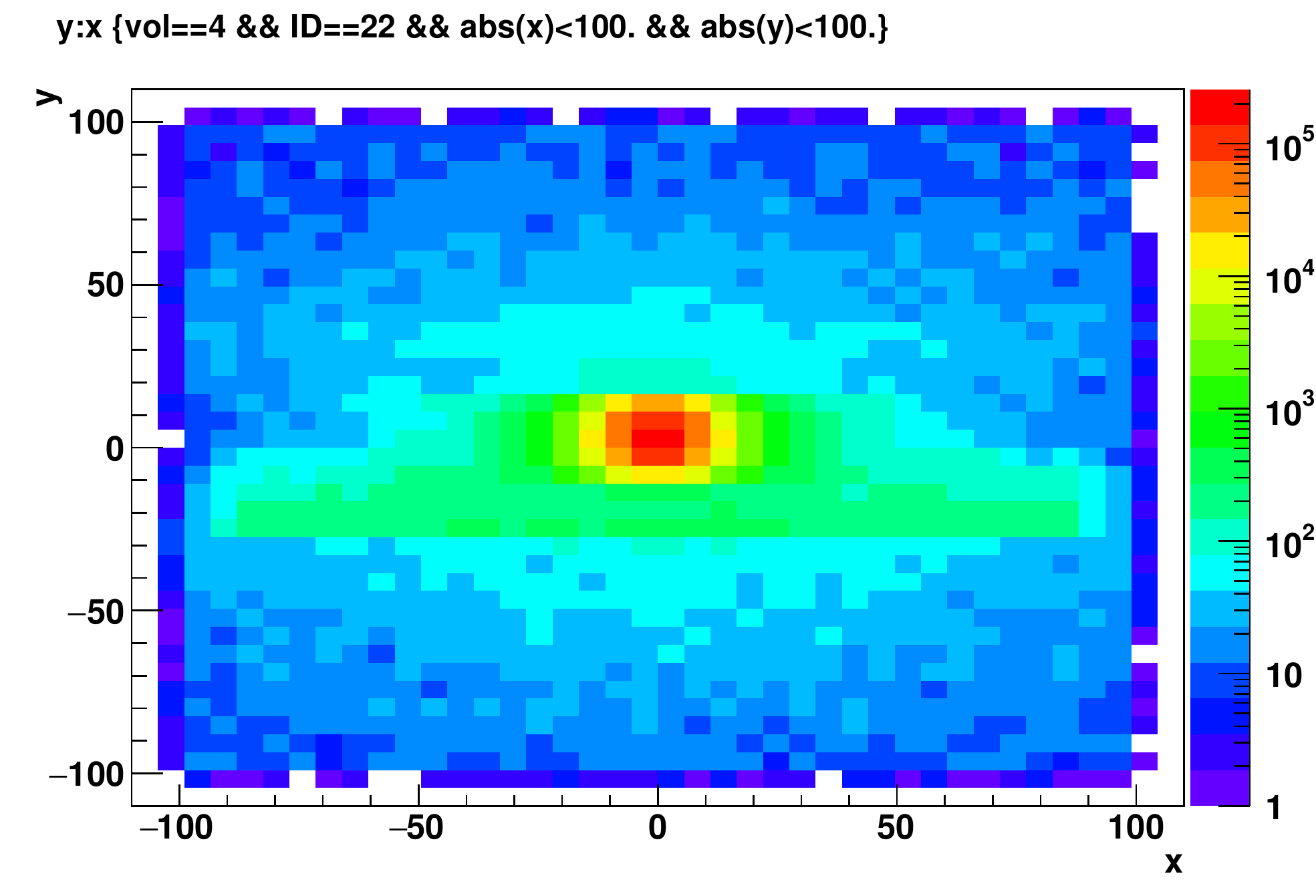} 
\end{center}
\caption{The second zoomed hit distribution in the forward direction at a distance of 15 m from the radiator
for photon energies above 1 keV. The coordinates X and Y are in millimeters.}
\label{fig:hits15c}
\end{figure}

\begin{figure}[!htbp]
\begin{center}
\includegraphics[trim = 0mm 0mm 0mm 0mm, width = 0.8\textwidth]
{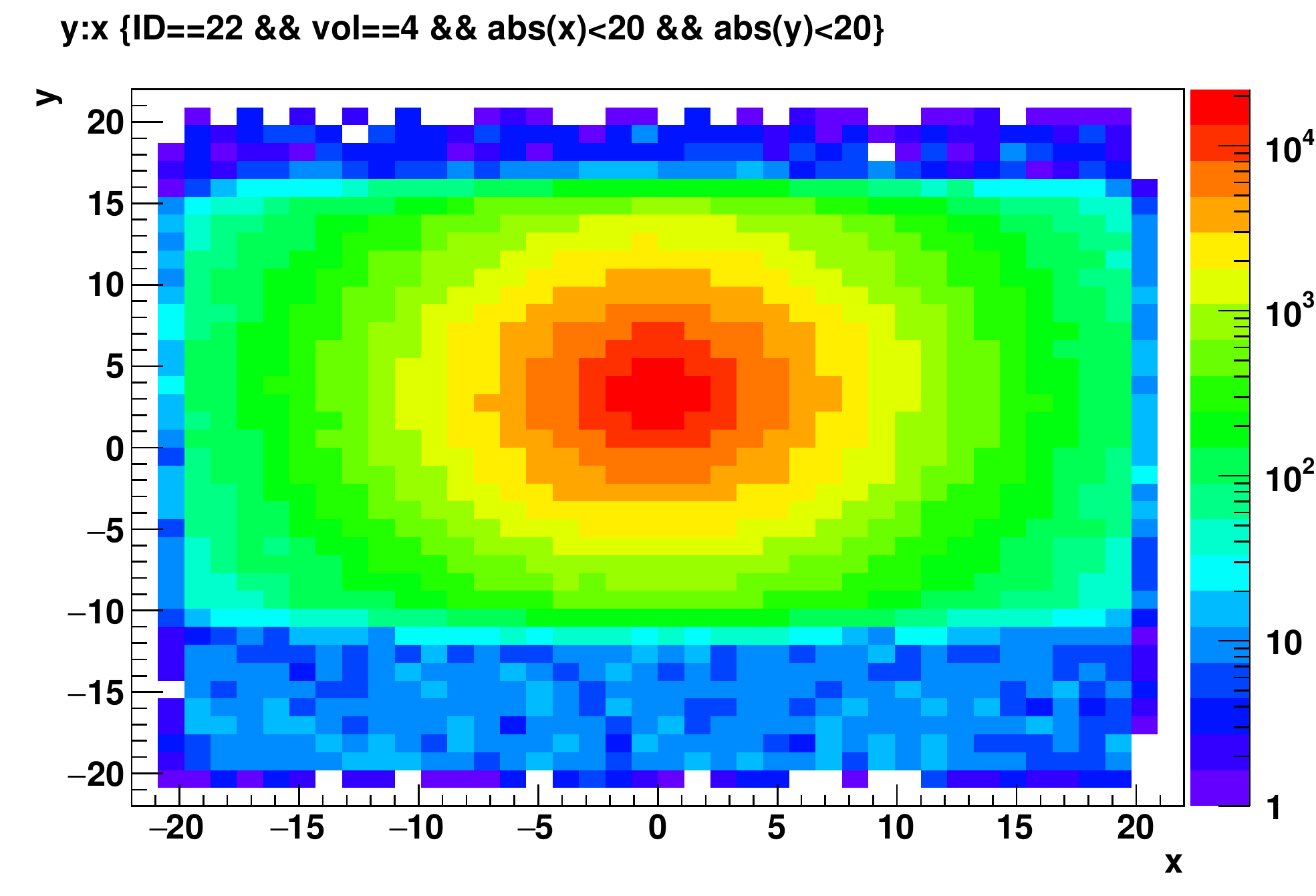} 
\end{center}
\caption{The third zoomed hit distribution in the forward direction at a distance of 15 m from the radiator
for photon energies above 1 keV. The coordinates X and Y are in millimeters.}
\label{fig:hits15f}
\end{figure}

Figure~\ref{fig:hits15a} shows the hit distribution for photons of all energies.
\begin{figure}[!htbp]
\begin{center}
\includegraphics[trim = 0mm 0mm 0mm 0mm, width = 0.8\textwidth]
{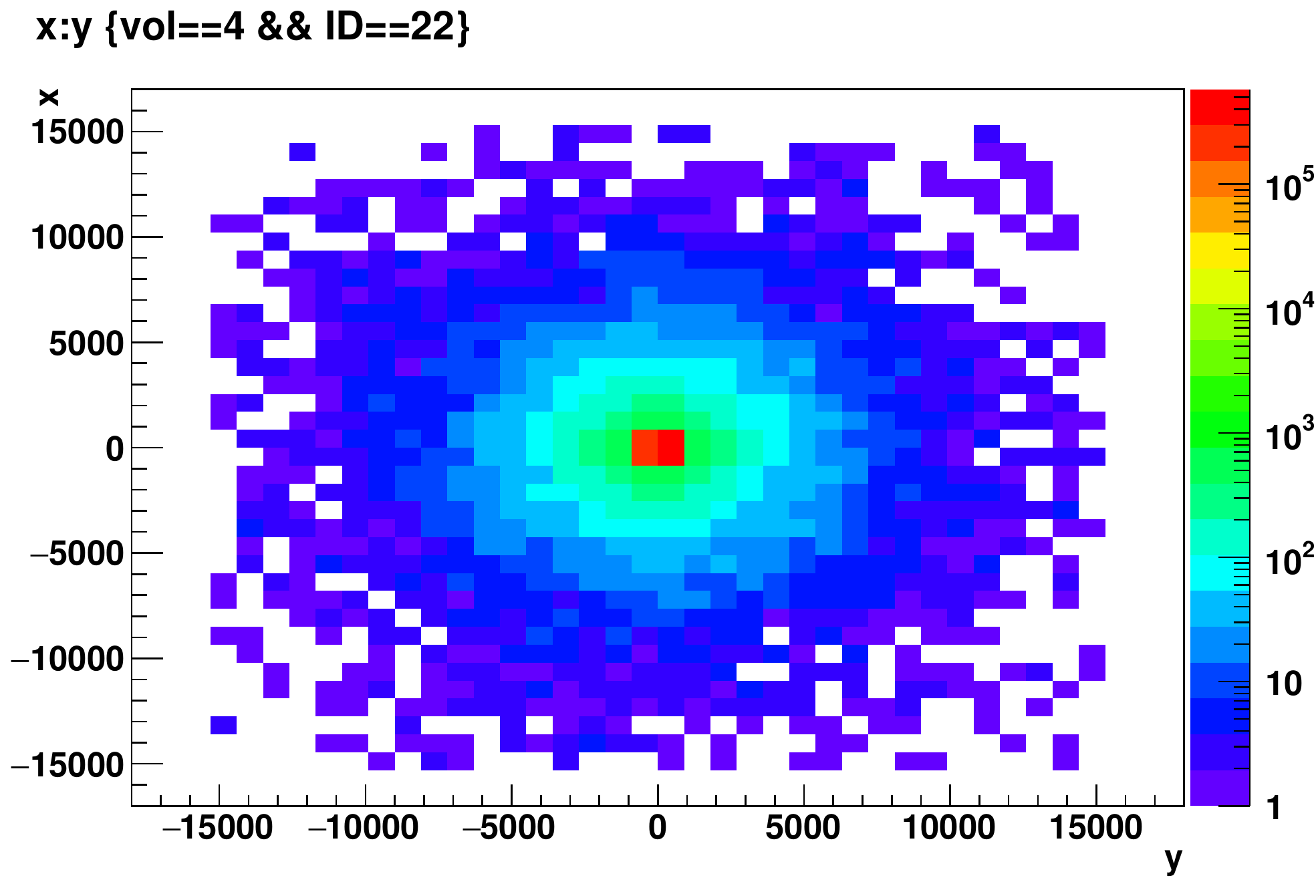} 
\end{center}
\caption{The hit distribution in the forward direction at a distance of 15 m from the radiator
for photon energies above 1 keV. The coordinates X and Y are in millimeters.}
\label{fig:hits15a}
\end{figure}
Figure~\ref{fig:hits15b} shows the same hit distribution in an area 1 meter by 1 meter.
Figure~\ref{fig:hits15c} shows the hit distribution in an area 20 cm by 20 cm.
Figure~\ref{fig:hits15f} shows the same in an area 4 cm by 4 cm (the footprint of the slots is recognizable).

\begin{figure}[!htbp]
\begin{center}
\includegraphics[trim = 0mm 0mm 0mm 0mm, width = 0.8\textwidth]
{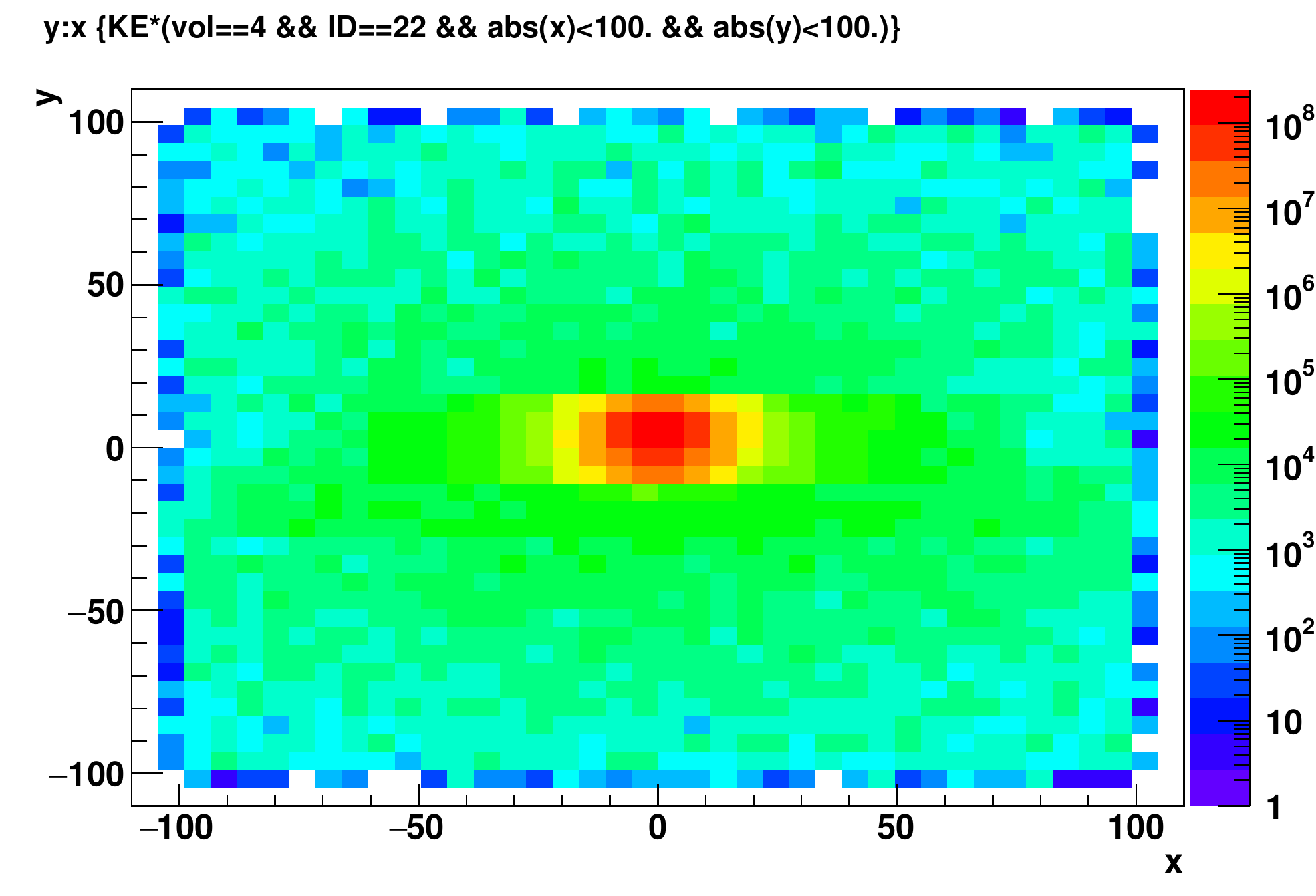} 
\end{center}
\caption{The energy flow distribution in the forward direction in an area 20 cm by 20 cm
at a distance of 15 m from the radiator. The coordinates X and Y are in millimeters.}
\label{fig:hits15d}
\end{figure}

\begin{figure}[!htbp]
\begin{center}
\includegraphics[trim = 0mm 0mm 0mm 0mm, width = 0.8\textwidth]
{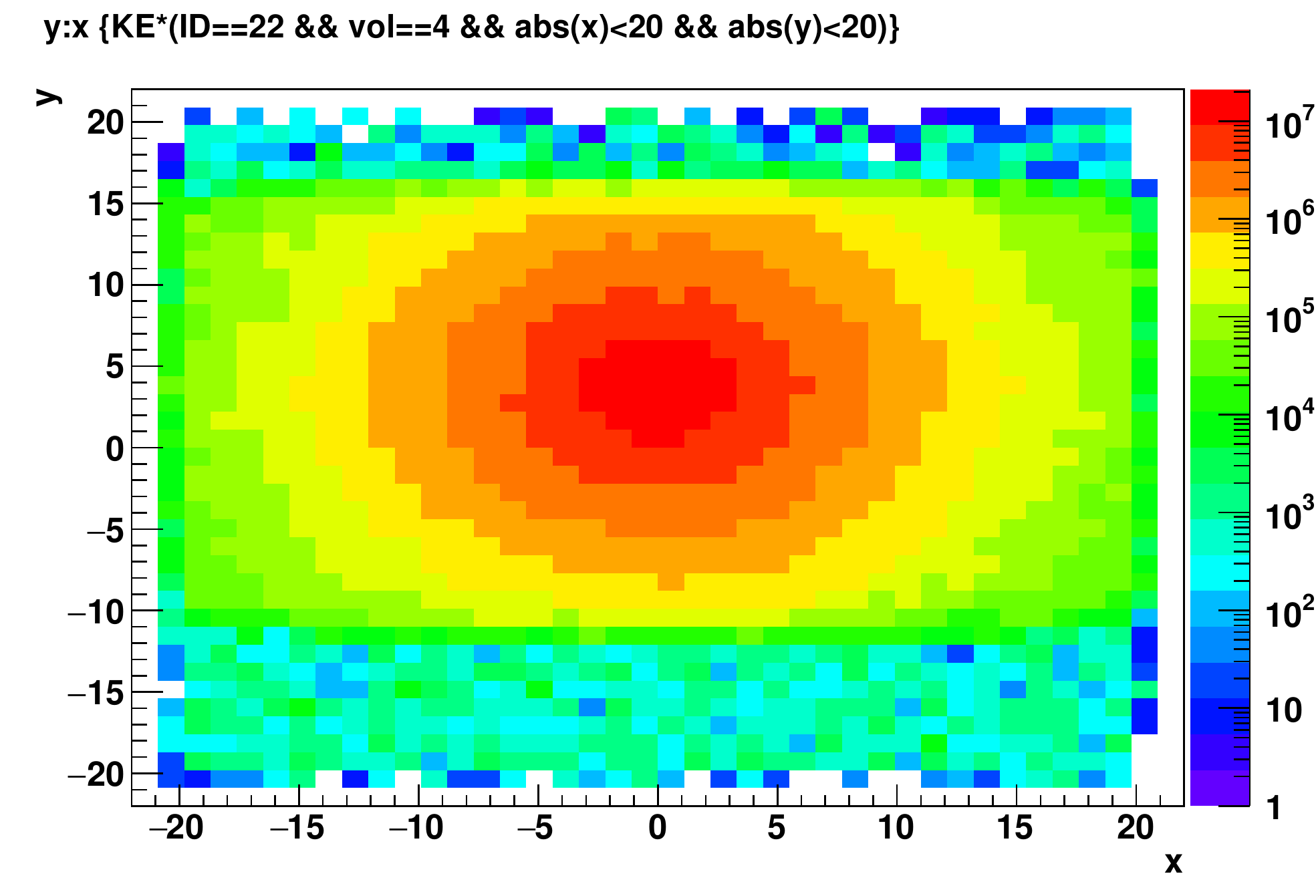} 
\end{center}
\caption{The energy flow distribution in the forward direction in an area 4 cm by 4 cm 
at a distance of 15 m from the radiator. The coordinates X and Y are in millimeters.}
\label{fig:hits15e}
\end{figure}

The energy flow from these photons is shown in 
Fig.~\ref{fig:hits15d}, which shows the distribution in an area 20 cm by 20 cm.
Fig.~\ref{fig:hits15e} shows the distribution in an area 4 cm by 4 cm.


\subsubsection {Radiation at the polarized target}

The radiation dose rate was found to be 3.9 rem/hour on the coil of the polarized target solenoid.
The corresponding integral over the experiment duration is 1 kRad, which presents
no problem for the equipment operation.

The intense photon beam leads to some power deposition in the low temperature
cell as well as inducing radiation defects in the material. These radical molecules,
at sufficient concentration, reduce the efficiency of the proton polarizing mechanism.

\paragraph {Heat load on the polarized cell}

\begin{figure}[h]
\begin{center}
\includegraphics[trim = 0mm 0mm 0mm 0mm, width = 0.6\textwidth]
{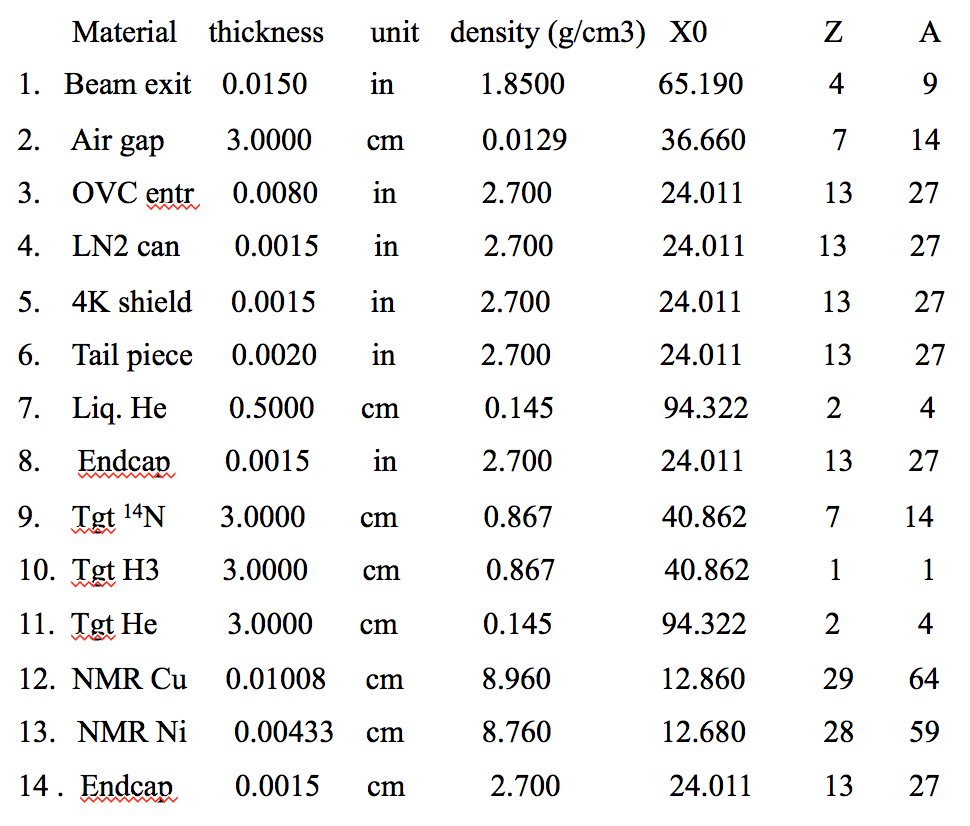} 
\end{center}
\caption{The list of materials for the NH$3$ target from Ref.~\cite{Donal}.}
\label{fig:materials}
\end{figure}

The full list of materials in the polarized target shown in Fig.~\ref{fig:materials} from Ref.~\cite{Donal} 
was implemented in the GEANT4 model.
The energy deposition obtained from MC normalized to the incident beam
intensity of 1.2 $\mu$A was found to be 190 mW.
The same power would be deposited by an electron beam of 35~nA.

\paragraph {Radiation damage to the NH$_3$ material} 

The radiation damage to the material is due to the creation of radical molecules, e.g. $NH_3e$.
These molecules have a very long lifetime at the low temperature of the polarized target. 
Under the beam irradiation they accumulate and compromise the rate of the proton polarizing process.
Annealing of the polarized material needs to be performed (warming up to \mbox{77$^\circ$ K}) after 
$10 \times 10^{15}$/cm$^2$ electrons have passed through the cold target~\cite{PolTarget}.
The production rate of radicals by a photon beam as well as by an electron beam
is proportional to the head load induced by the ionization of the same material.
Using the result of the heat load analysis (see the paragraph above), the required 
rate of annealing cycles was found to be every 40 hours.

\subsection{Photon distribution in the forward direction}
\begin{figure}[!htbp]
\begin{minipage}[h]{0.75\textwidth}
\begin{center}
\includegraphics[trim = 0mm 0mm 0mm 0mm, width = 0.62\textwidth]{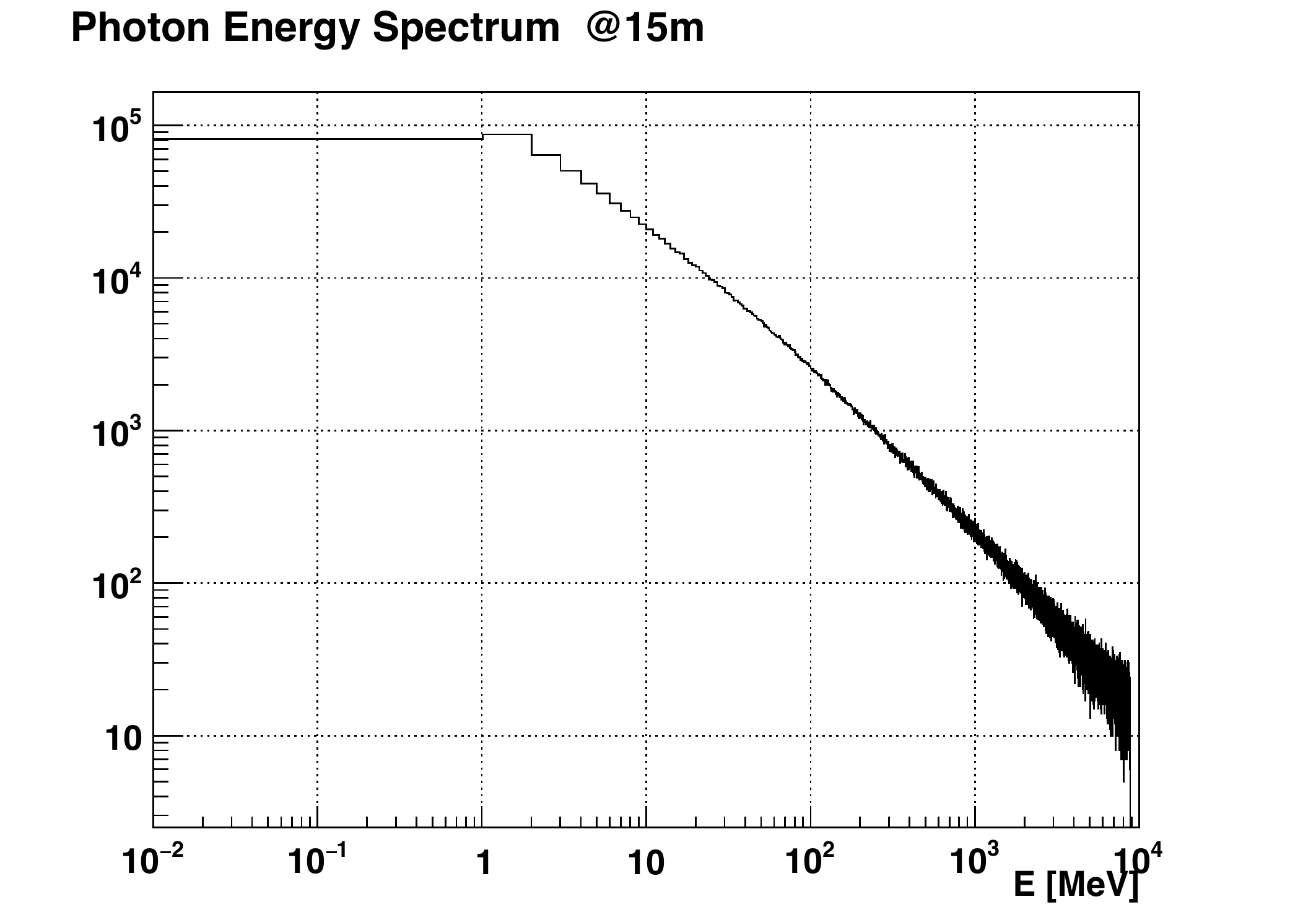} 
\end{center}
\end{minipage}
\begin{minipage}[h]{0.20\textwidth}
\begin{center}
\caption{The photon energy spectrum in the logarithmic scale.}
\label{fig:energies1}
\end{center}
\end{minipage}
\end{figure}

\begin{figure}[!htbp]
\begin{minipage}[h]{0.75\textwidth}
\begin{center}
\includegraphics[trim = 0mm 0mm 0mm 0mm, width = 0.62\textwidth]{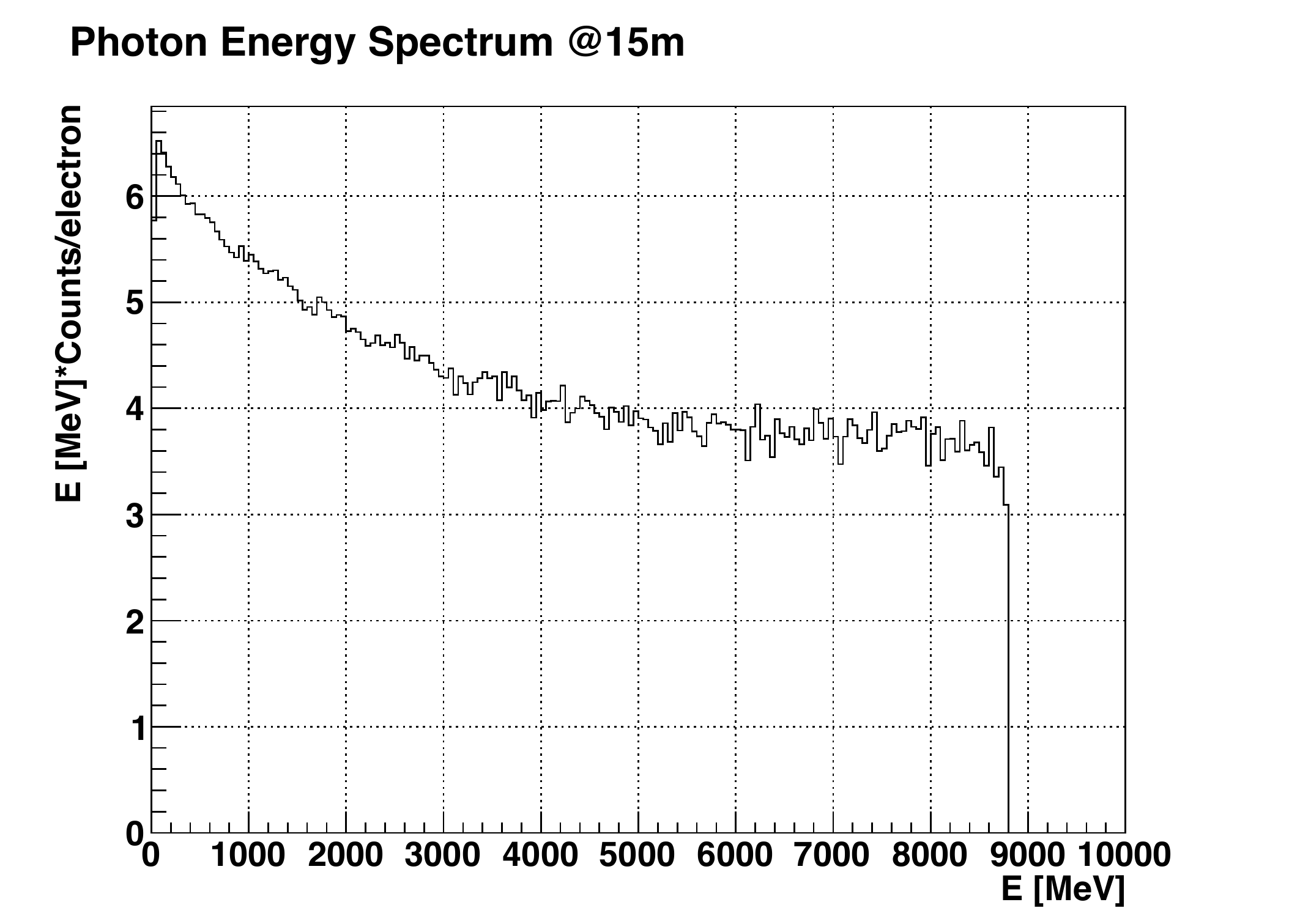} 
\end{center}
\end{minipage}
\begin{minipage}[h]{0.20\textwidth}
\begin{center}
\caption{The photon energy spectrum in the linear scale.}
\label{fig:energies2}
\end{center}
\end{minipage}
\end{figure}

\begin{figure}[!htbp]
\begin{minipage}[h]{0.75\textwidth}
\begin{center}
\includegraphics[trim = 0mm 0mm 0mm 0mm, width = 0.65\textwidth]{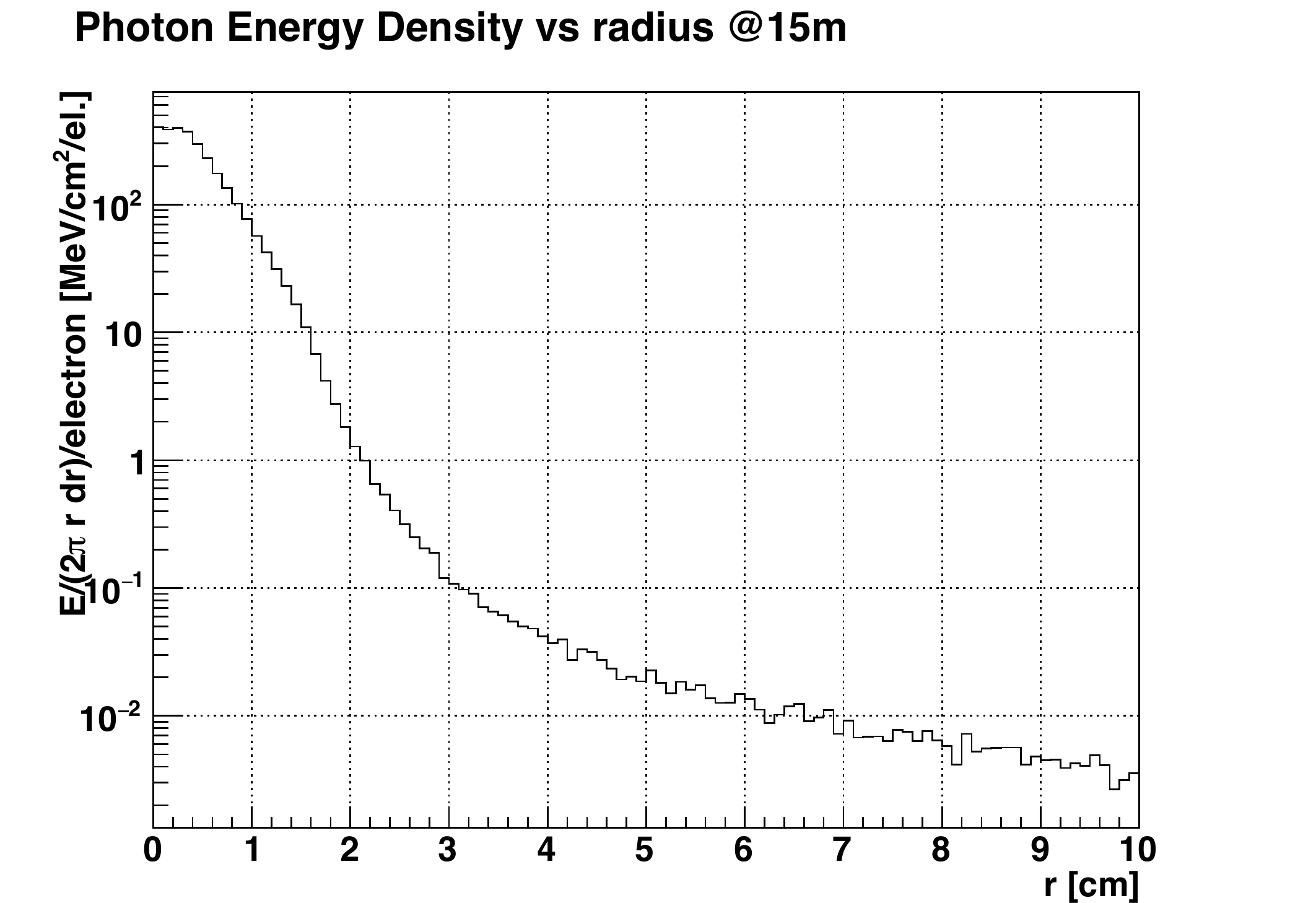} 
\end{center}
\end{minipage}
\begin{minipage}[h]{0.20\textwidth}
\begin{center}
\caption{The photon energy density at 15 meters from the radiator.}
\label{fig:power1}
\end{center}
\end{minipage}
\end{figure}

\begin{figure}[!htbp]
\begin{minipage}[h]{0.75\textwidth}
\begin{center}
\includegraphics[trim = 0mm 0mm 0mm 0mm, width = 0.65\textwidth]{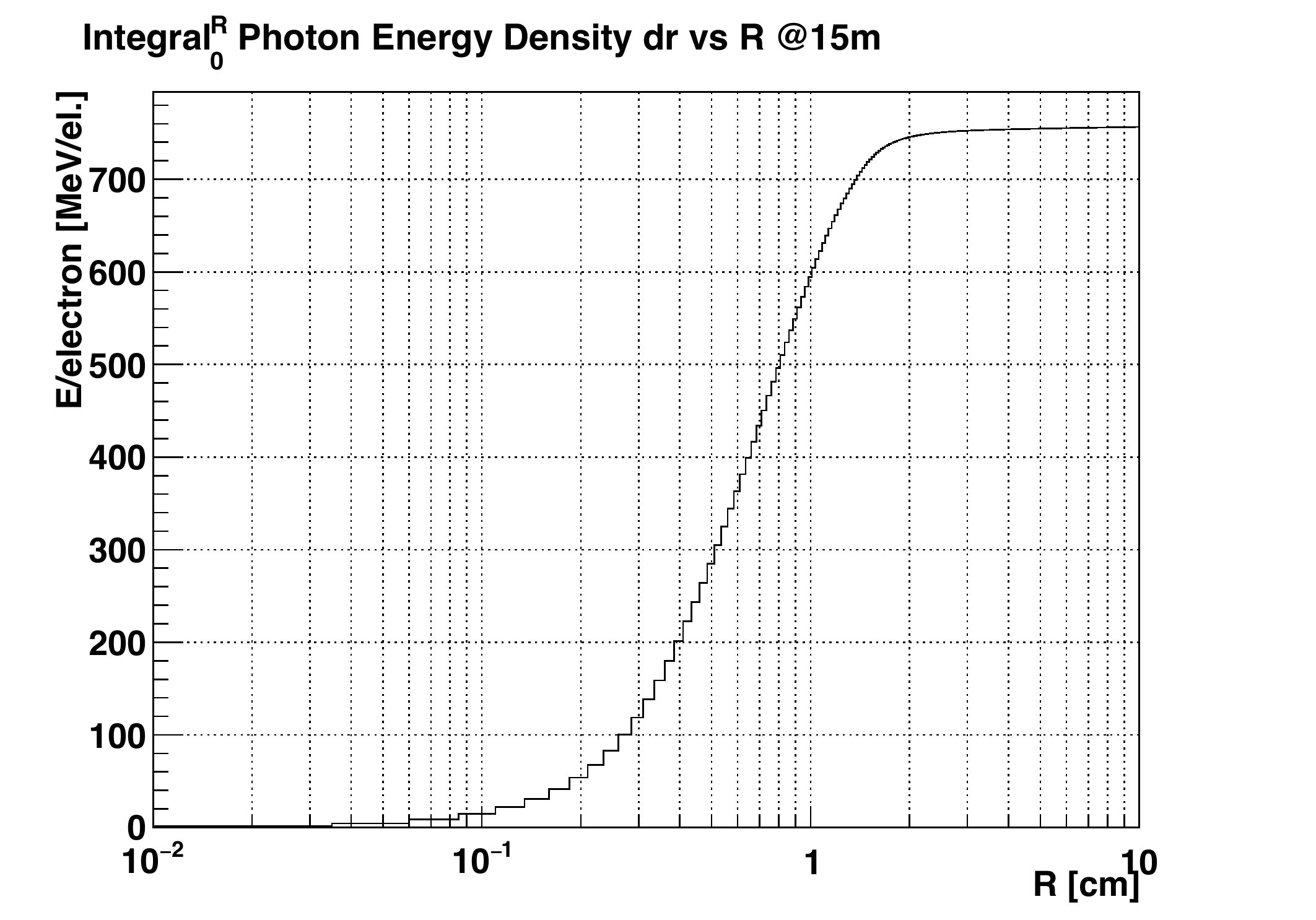} 
\end{center}
\end{minipage}
\begin{minipage}[h]{0.20\textwidth}
\begin{center}
\caption{The photon energy flow integral.}
\label{fig:power2}
\end{center}
\end{minipage}
\end{figure}

\ind The energy and coordinate distribution of the photons were analyzed at a distance of
15 meters from the radiator.
Figures~\ref{fig:energies1} and  \ref{fig:energies2} show the energy spectrum 
in the logarithmic and linear scales.
Figures~\ref{fig:power1} and \ref{fig:power2} show the radial distribution of the energy flow and its integral.
The total energy of the photons in the forward direction (integrated up to a 15 meter distance from the beam)
is 770 MeV per incident electron (or 0.92 kW for the 1.2~$\mu$A beam intensity),
which is close to the expected value for a 10\% radiation length radiator and 8.8~GeV beam energy.

An important observation made from this MC study is that about 99\% of the total energy flow of photons
is concentrated inside a 2-cm radius circle.
The beam pipe to the Hall A beam dump and the beam dump itself have apertures for the $\pm$0.5 
degree cone which, at a distance of 15~meters from the radiator, translates to a radius of 13~cm.
The total power of photons outside a radius 10~cm was found to be below 2.5 W.
This means that almost all photon power is going to the main Hall A beam dump.

\section{Radiation in the hall induced by the photon beam on the polarized target}

\ind In the presence of the polarized cell in the beam line, a significant fraction of photons (3.7\%) will
interact with the target material and convert into electron-positron pairs as well as producing hadrons.
Those pairs are also very well aligned with the direction of incident photons.
However, after they leave the solenoid, their trajectories deflect from the beam line direction. 
Some part of that power (less than 35 W) will contribute to the radiation level in the hall. 

The radiation level in the hall will have a contribution from the particle
production by the photon beam on the polarized target.
Instead of a direct calculation of such a background, 
we will compare the $\gamma$-nucleon luminosity of the proposed setup with 
the effective $\gamma$-nucleon luminosity of the GEP experiment 
which is based on the same SBS spectrometer.
The proposed WACS experiment luminosity is about $10^{36}$ Hz/cm$^2$ equivalent gamma-nucleon.
The GEP experiment electron-nucleon luminosity is $8\times10^{38}$ Hz/cm$^2$  with a 40-cm long
LH2 target, which is equivalent to the luminosity $4\times10^{37}$ Hz/cm$^2$ equivalent gamma-nucleon.
It is easy to see that background rates would be 40 times lower in the proposed experiment
than in the GEP experiment.

There is an interesting experimental observation which supports the consideration above: 
The 2006 GEN and 2008 d2n experiments with open geometry detectors in Hall A 
(the BigBite spectrometer and the neutron detectors) 
operated at electron-nucleon luminosity of $1 \times 10^{37}$ Hz/cm$^2$, 
or at the luminosity $0.5\times10^{36}$ Hz/cm$^2$ equivalent gamma-nucleon, 
which is comparable to the equivalent gamma-nucleon luminosity we plan to use in 
the proposed setup ($10^{36}$ Hz/cm$^2$ equivalent gamma-nucleon).
At that time, the BigBite tracker used drift chambers, whose hit rate capability was much lower 
than the hit rate at which the GEM chamber of SBS can operate.

\section{TOSCA calculations of the field}

\ind There are two aspects of this device which are concerned with the magnetic field outside of the dipole.
They are the field gradient in the polarized cell region in the proposed configuration
and the force between the solenoid of the polarized target and the iron structure of the magnet-dump.

\begin{figure}[!htbp]
\begin{center}
\includegraphics[trim = 0mm 0mm 0mm 0mm, width = 0.65\textwidth]{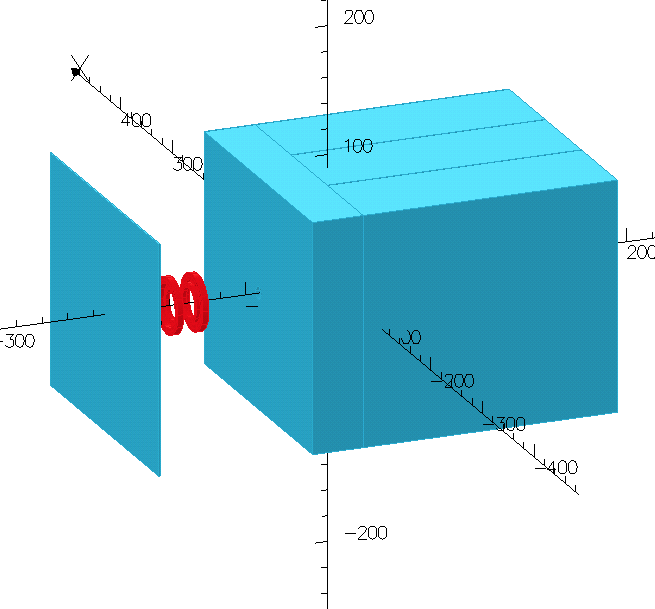} 
\end{center}
\caption{The view of the magnet-dump shielding with the polarized target coils and the compensating plates
(left to right).}
\label{fig:compensator}
\end{figure}

The relative field gradient was found to be below $10^{-4}$, which is sufficiently low for the target
operation. 
The calculated force in the z direction on all coils was found to be 500~kG 
(five times above the recommended limit of 100 kG~\cite{Dave}).
To resolve this issue we added a 5-cm thick steel compensation plate, see Fig.~\ref{fig:compensator},
after which the force became below 20~kG.
In the actual experiment there are additional blocks of iron in similar proximity to the target.
They are the SBS spectrometer magnet and its support structure.
This suggests that the enclosure of the target in a four-sided steel cage would be a more reliable scheme.
It would also help to control the force in the x direction.

\section{The plan for further testing and development}

\ind There are several improvements of the design which could be made, for example
a step profile of the narrow channels in the CuW80 insert (for reduction of the photon
leakage) and non-magnetic shielding material (for reduction of the force between
the source and the solenoid). 

The analysis presented in this document could be be confirmed by 
the measurement of the radiation level and detector counting rates with a very low intensity 
beam of high energy in a thick target surrounded by the proposed shielding.
The target would be made of a heavy metal bar with a channel oriented at one degree
with respect to the beam direction to imitate the geometry of the shower in the case
of the magnetic field, so no magnet would be used in such a test.

\bibliographystyle{plain}

\begin{thebibliography}{99}
\bibitem{WACS-ALL} S.~Abrahamyan, G.~Niculescu, B.~Wojtsekhowski, Proposal PR12-15-003 to PAC43.
\bibitem {Moskov} M.~Amaryan and P.~Degtiarenko, private communication, 2015.
\bibitem{PDG}  K.A. Olive \etal (Particle Data Group), {\it Chin. Phys.} {\bf C}, 38, 090001 (2014).
\bibitem{Swanson} W.P. Swanson, SLAC-PUB 2042, 1977.
\bibitem{gdml} \url{http://gdml.web.cern.ch/GDML/doc/GDMLmanual.pdf}
\bibitem {ranecu} F.~James, Comp. {\em Phys. Comm.} {\bf 60}, 329-344 (1990).
\bibitem {bertini} M.P.~Guthrie, R.G.~Alsmiller and H.W.~Bertini, \NIM A {\bf 66}, 29 (1968); 
H.W.~Bertini and M.P.~Guthrie, {\em Nucl. Phys.~A} {\bf 169}, (1971).
\bibitem {fritiof} B.~Andersson \etal, {\em Nucl. Phys. B } {\bf 281} 289 (1987); 
B.~Nilsson-Almquist, E.~Stenlund, {\em Comp. Phys. Comm. } {\bf 43} 387 (1987).
\bibitem {g4paper} Geant4 collaboration, Geant4 general paper (to be published), \NIM A, (2003).
\bibitem {Pavel} P.~Degtiarenko, private communication, 2001.
\bibitem{Kapton} V.V.~Petrov, Yu.A.~Pupkov,  a report ``BINP TESTING OF RADIATION RESISTANCE
OF THE MATERIALS USED FOR PRODUCTION OF ACCELERATOR MAGNETIC SYSTEMS", 
Novosibirsk, 2011.
\bibitem{Donal} D.~Day, private communication, 2015.
\bibitem{PolTarget} J.~Pierce \etal, NIM A 738 (2014) 54.
\bibitem{Dave} D.~Meekins, private communication, 2015.
%
\end{thebibliography}

\end{document}